\documentclass[11pt,a4paper]{article}

\usepackage[a4paper, total={6in, 8in}]{geometry}
\usepackage{authblk}

\usepackage[utf8]{inputenc}
\usepackage{subcaption}    
\usepackage{hyperref}
\usepackage{booktabs}
\usepackage{%
	amsfonts,%
	amsmath,%
	float,%
	etoolbox,%
	mathrsfs,%
	mathtools,%
	multirow,%
	pgf,%
	pgfplots,%
	tikz%
}

\usepackage{amssymb}

\usepackage{todonotes}
\usepackage{siunitx}
\usepackage{algorithm}
\usepackage{algorithmic}
\newtheorem{rem}{Remark}
\usepackage{bm}
\usepackage{graphicx}

\title{Adaptive Aborting Schemes for Quantum Error
Correction Decoding}
\author[$*$]{Sanidhay Bhambay}
\author[$\dag$]{Prakash Murali}
\author[$*$]{Neil Walton}
\author[$*$]{Thirupathaiah Vasantam}
\affil[$*$]{Durham University, UK}
\affil[$\dag$]{Cambridge University, UK}

\date{}

\begin{document}
\maketitle
\begin{abstract}
Quantum error correction (QEC) is essential for realizing fault‐tolerant quantum computation. Current QEC controllers execute all scheduled syndrome (parity bit) measurement rounds before decoding, even when early syndrome data indicates that the run will result in an error.
The resulting excess measurements increase the decoder’s workload and system latency. To circumvent this, 
we introduce an \textit{adaptive-abort} module which simultaneously improves the decoder overhead and suppresses the logical error rates of surface codes and color codes with the existing QEC controller.
The key idea is that initial syndrome information allows the controller to terminate risky shots early before additional resources are spent. An effective scheme balances the cost of further measurement against the restart cost and thus increases decoder efficiency.

Adaptive abort schemes dynamically adjust the number of syndrome measurement rounds per shot using real‐time syndrome information. We considered three schemes: Fixed‐depth (FD) decoding, the standard non-adaptive approach used in current state-of-the-art QEC controllers, and two adaptive schemes \textit{AdAbort} and \textit{One‐Step Lookahead} (OSLA) decoding. For the surface and color codes under a realistic circuit‐level depolarizing noise model, AdAbort substantially outperforms both OSLA and FD, yielding higher \textit{decoder efficiency}  across a broad range of code distances. 
Numerically, as the code distance increases from $5$ and $15$, AdAbort yields an improvement increasing from $5 \%$ to $35\%$ for surface codes, and $7 \%$ to $60 \%$ for color codes.

To our knowledge, these are the first adaptive abort schemes to be considered for QEC. Our results highlight the potential importance of abort rules in increasing efficiency as we scale to large resource-intensive quantum architectures.
\end{abstract}

\section{Introduction}

Quantum computers have the potential to solve problems~\cite{cao2019quantum,peruzzo2014variational,scarani2009security} beyond the reach of today’s supercomputers. 
However, hardware for quantum computers is extremely fragile: qubits (quantum bits) are highly sensitive, and even small interactions with their surroundings can erase quantum information through decoherence~\cite{bravyi2024high}. Moreover, every logical operation (that is, a gate applied to an encoded qubit) incurs significant overhead because it must be protected with many repeated rounds of syndrome measurements and classical decoding~\cite{fowler2011two,gottesman2009introduction}. 
Therefore, these syndrome measurement rounds increase both the amount of syndrome data the decoder must process and the number of times the decoder must run—this significantly increases the decoder’s workload and lowers decoder efficiency, defined as the number of correctly decoded logical outputs produced per unit execution time.
Hence, larger quantum algorithms not only increase the chance of decoherence but also consume more scarce decoder and measurement resources. To address this, we explore adaptive early-abort schemes that abort hopeless runs before they reach the decoder, thereby reducing the burden on the decoder and improving decoder efficiency.

Quantum Error Correction (QEC)~\cite{shor1995scheme,steane1996error,terhal2015quantum,nielsen2010quantum} is critical to keep quantum information alive long enough for useful quantum computation.  Instead of storing a single logical qubit in one physical qubit, QEC encodes it redundantly across many physical qubits, in the same way that classical error-correcting codes (ECC) redundantly encode a logical bit across multiple physical bits. This way, QEC can detect and correct errors on individual physical qubits. 
Similar to classical ECC, for QEC to function, the hardware must support an error rate below a certain \emph{logical qubit threshold} or \emph{threshold}~\cite{gottesman2009introduction,preskill1998reliable,aharonov1997fault,preskill1998reliable}, ensuring that, even if several physical qubits undergo errors, the original logical state can still be recovered.

Leading QEC schemes like the surface code~\cite{kitaev2003fault,fowler2012surface}, color code~\cite{bombin2006topological,fowler2011two,landahl2011fault} and the stabilizer codes~\cite{shor1995scheme}, 
distribute quantum information in a way that errors manifest as detectable syndromes or parity bits (see Figure~\ref{fig:measuring_ckt}).
When these syndromes are combined with real-time classical decoding systems, QEC can suppress logical error rates exponentially below physical error rates.
This enables fault-tolerant quantum computation (FTQC), where errors are corrected faster than they accumulate, allowing arbitrarily long quantum computations~\cite{nielsen2010quantum,preskill1998reliable}.
Applications such as simulating complex molecules for drug discovery~\cite{cao2019quantum}, material design~\cite{peruzzo2014variational}, cryptography~\cite{scarani2009security,bhambay2025proportional2}, and large-scale optimization~\cite{fakhimi2023quantum} problems require billions of error-corrected quantum operations~\cite{gidney2021factor,beverland2022assessing}, which is far beyond the reach of today's noisy intermediate-scale quantum (NISQ) devices. As a result, industry leaders such as IBM~\cite{smith-goodson2025ibm}, Google~\cite{Googlemilestone}, and Microsoft~\cite{Microsoftmajorana1} have made QEC central to their roadmaps. 

There is increasing recognition that decoders must not only minimize logical error rates but also operate at a throughput sufficient to keep pace with large-scale quantum computations. As code distances and system sizes grow, the decoder can become a bottleneck in the overall architecture. Efforts to address this include parallelization \cite{fowler2012towards,tan2023scalable} and specialized hardware implementations \cite{ziad2025local}; here, we instead reduce decoder workload with minimal system overhead by supplying it primarily with executions that likely to succeed in decoding.
Emphasize that low-overhead decoding is important.

In a fault‑tolerant quantum computer, every logical operation is surrounded by many consecutive rounds of syndrome measurement that detect and correct errors in real time.
For instance, implementing a logical $T$-gate or a logical CNOT on a distance $5$ surface code requires a few hundred syndrome measurement rounds to suppress the logical error rate below an acceptable threshold~\cite{fowler2012surface}.
Furthermore, because logical qubits still incur errors, each quantum algorithm must be executed multiple times in separate shots, with each shot being a full end‑to‑end execution of the algorithm interleaved with syndrome measurement and correction. 
If shots include many syndrome measurement rounds, each shot generates more syndrome data and forces the decoder to run more often. This increases the decoder workload and reduces its efficiency defined as the number of successfully decoded logical outputs the system can deliver for a given measurement and decoding budget.

\begin{figure*}[t]            
  \centering
  \begin{subfigure}{0.49\textwidth}
    \includegraphics[width=\linewidth]{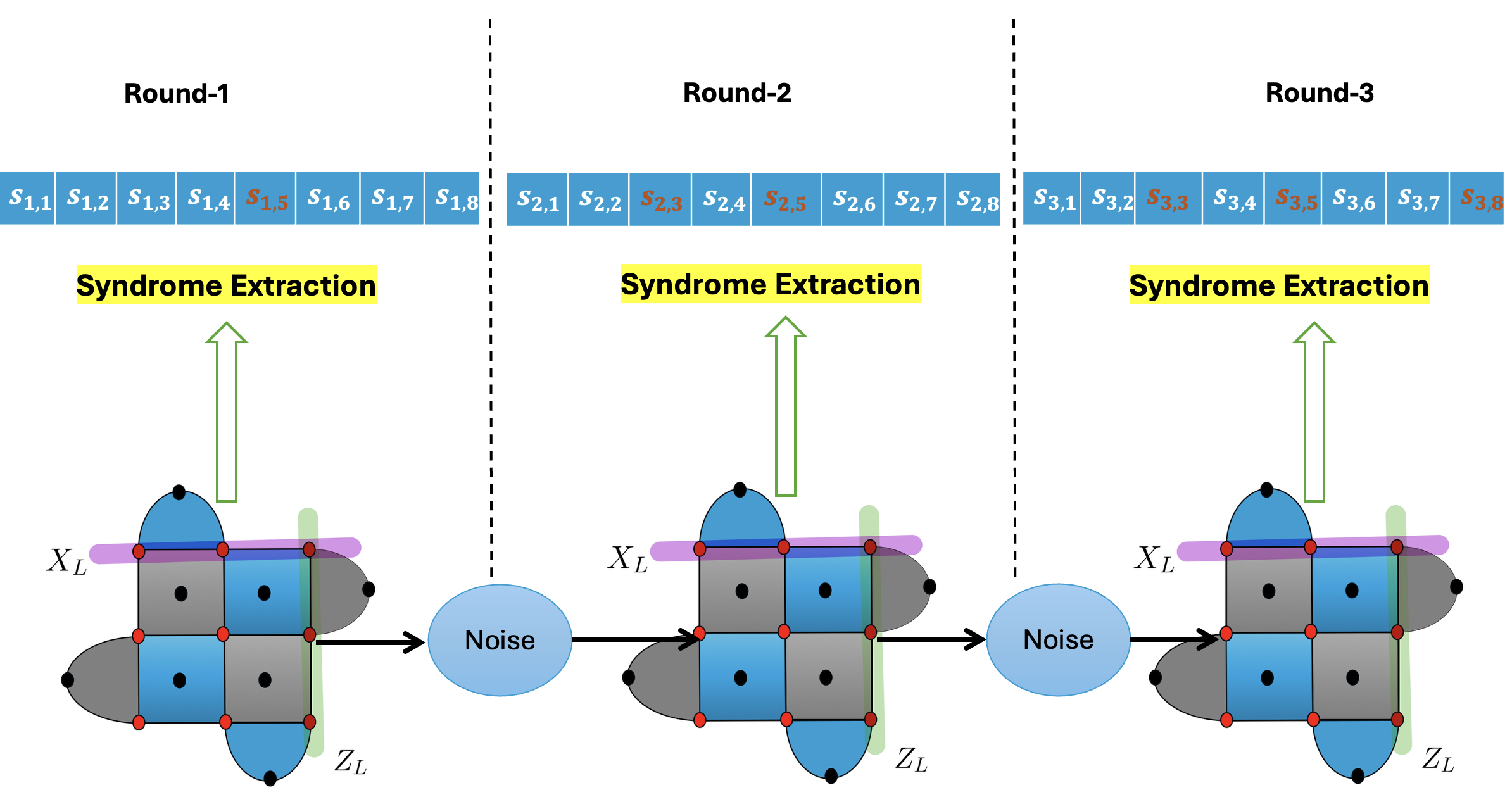}
    \caption{Shot representation for three syndrome‑measurement rounds using an FD decoding scheme}
    \label{fig:1a}
  \end{subfigure}\hfill
  \begin{subfigure}{0.43\textwidth}
    \includegraphics[width=\linewidth]{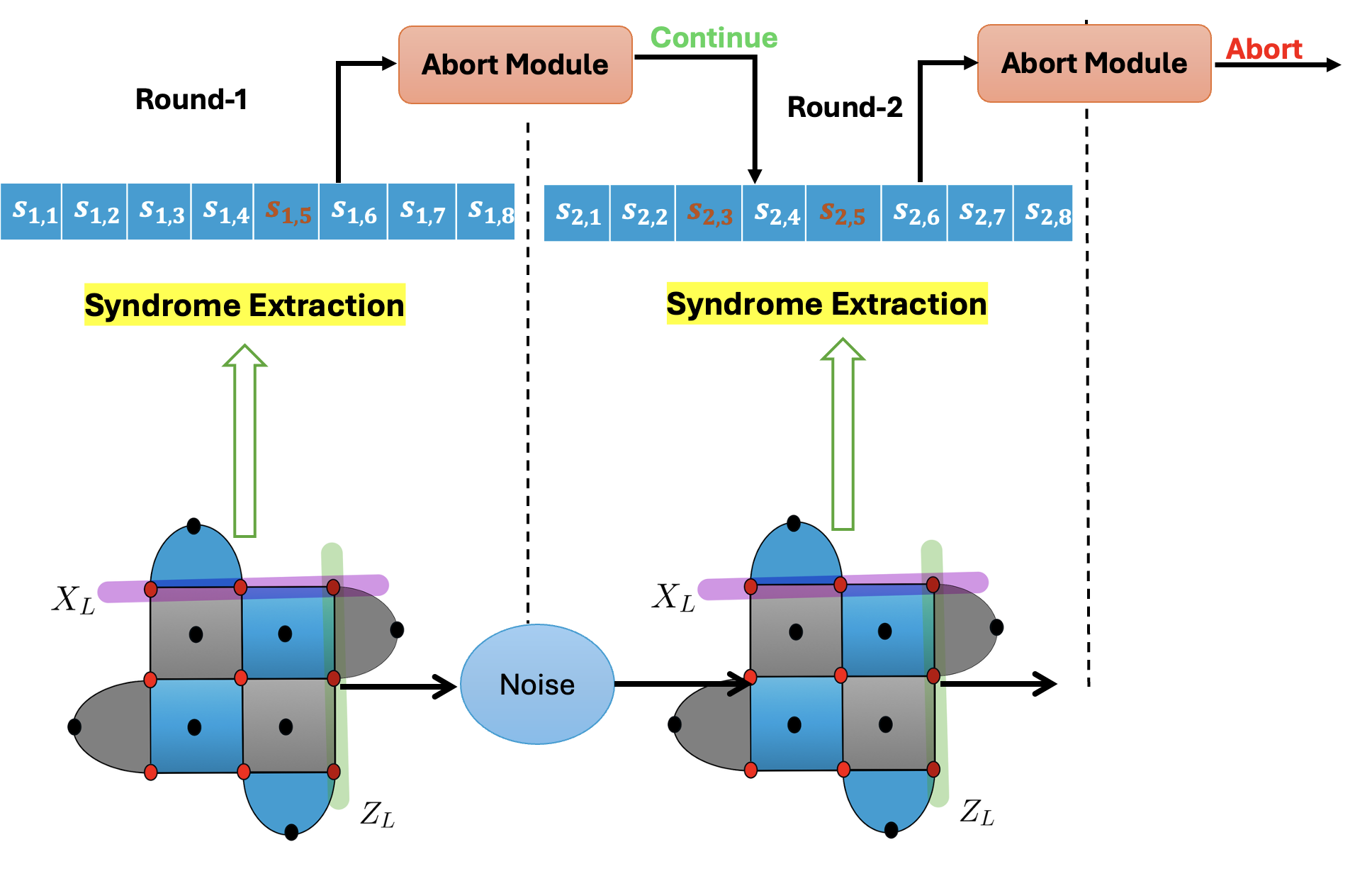}
    \caption{Shot representation for three syndrome‑measurement rounds using adaptive schemes along with the abort‑module.}
    \label{fig:abort_shot}
  \end{subfigure}
  \caption{Sub-figures (a) and (b) illustrate how adaptive schemes alter the number of measurement rounds compared to the FD scheme for the $[9,1,3]$ surface code. In both cases, $\mathbf{s}_i = [s_{i,1}, s_{i,2}, \dots, s_{i,8}]$ denotes the eight syndrome bits obtained from the $i$-th measurement round, with bits highlighted in red indicating detected errors. In (a), under the FD scheme, the circuit executes all three predetermined syndrome measurement rounds regardless of the observed syndromes. In (b), the adaptive scheme terminates the measurement process early, after the second round, based on the observed syndrome pattern, avoiding unnecessary additional measurements.}
  \label{fig:comparison_shots}
\end{figure*}

Figure~\ref{fig:comparison_shots} illustrates the typical structure of a shot with repeated syndrome measurement rounds.

\begin{figure}[htb]        
  \centering
  \begin{subfigure}{0.45\columnwidth}
    \includegraphics[width=\linewidth]{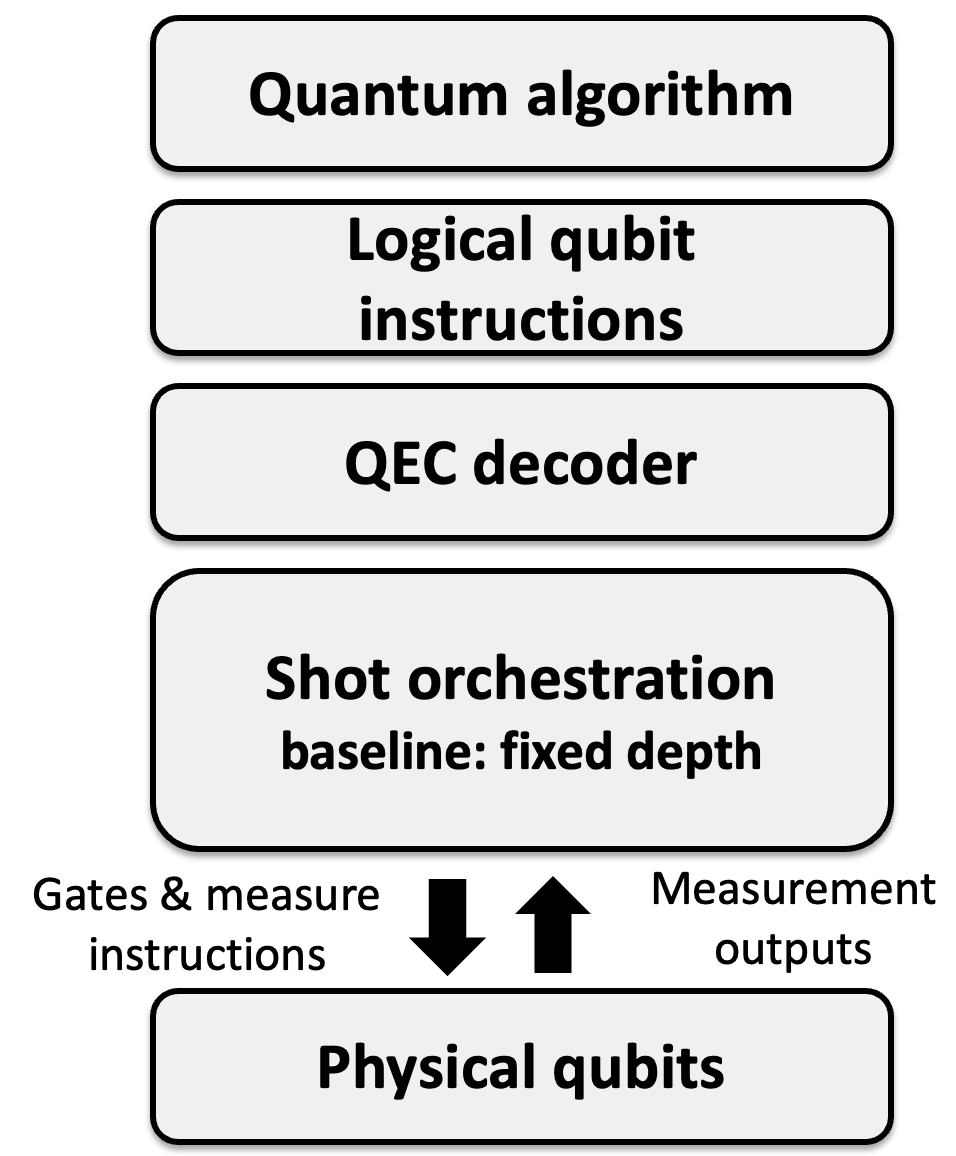}
    \caption{QEC stack architecture}
    \label{fig:fd_module}
  \end{subfigure}\hfill
  \begin{subfigure}{0.5\columnwidth}
    \includegraphics[width=\linewidth]{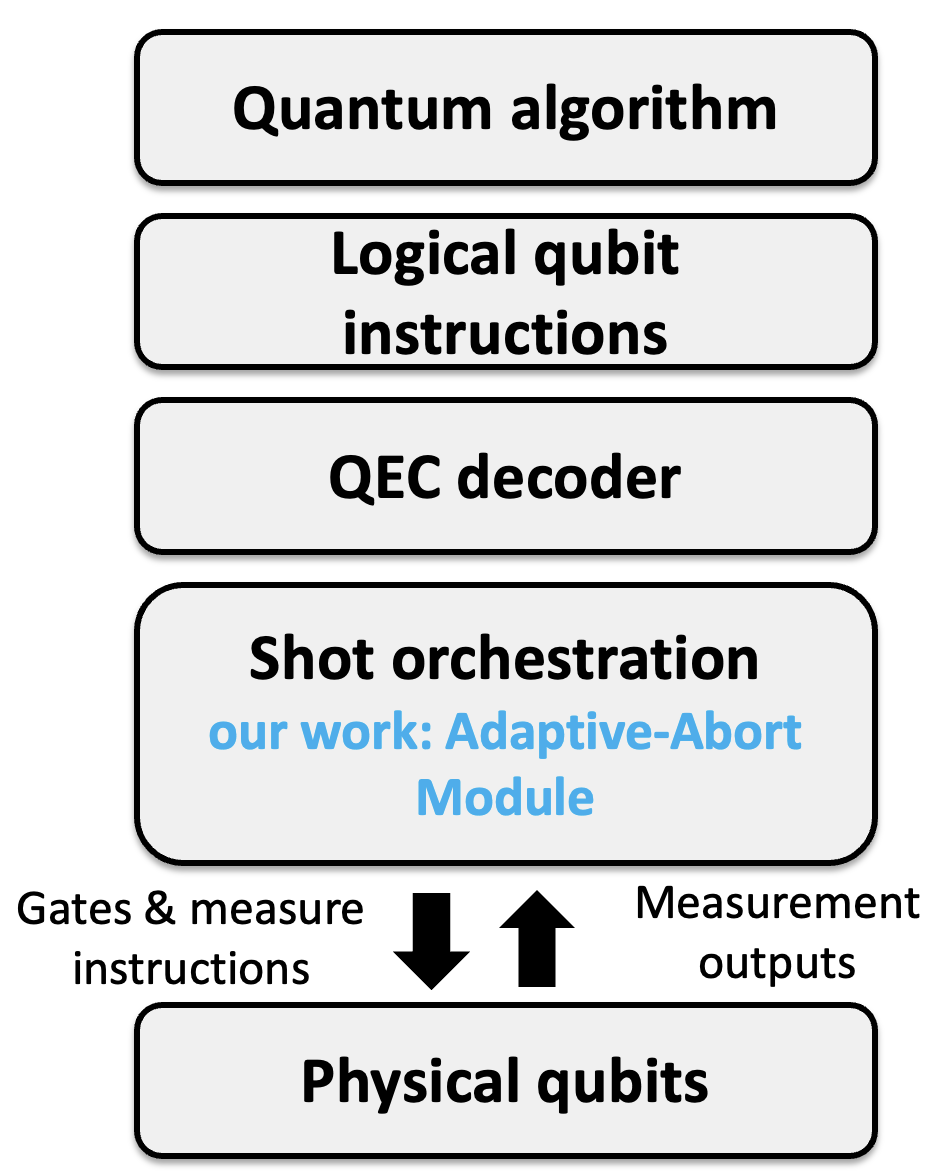}
    \caption{QEC stack with abort module}
    \label{fig:abort_module}
  \end{subfigure}
  \caption{a) Current state‑of‑the‑art QEC stack architecture, b) QEC controller architecture augmented with an adaptive abort module. The adaptive-abort module sits within the shot orchestration layer, between the QEC decoder and the physical qubits, continuously evaluating decoding confidence and issuing early aborts.}
  \label{fig:}
\end{figure}

Today’s state-of-the-art QEC controllers implement a fixed‐depth (FD) scheme where each shot always runs the predefined number of syndrome measurement rounds before decoding, with no mechanism to abort early.
Figure~\ref{fig:fd_module} shows the layered architecture, from the physical qubits at the bottom, through the shot orchestration layer and the QEC decoder, up to logical qubit instructions and the high-level quantum algorithm framework. In a conventional FD-only design, every shot must pay for all syndrome measurement rounds, regardless of its eventual success or failure, which increases the data processing at the decoder and its workload.
Rather than waiting until all syndrome measurements are completed, we view syndrome extraction as a sequential decision problem under noise. 
There is a growing body of work on adaptive syndrome measurement \cite{delfosse2020short,tansuwannont2023adaptive,berthusen2025adaptive}; however, this appears to be the first work to focus on controlled execution, throughput, and efficiency, rather than purely on logical error reduction. In particular, after each round, the controller has observed part of the syndrome and can already assess the likelihood of eventual decoding success. Since full executions and decoding operations are expensive, resources can in principle be saved by aborting an unpromising sequence of syndrome measurements early. We focus on adaptive abort decisions at the controller level. At each step, the system must decide whether there is sufficient evidence to justify continuing with additional syndrome measurements or whether it is preferable to accept the losses incurred so far and restart the process. Our insight is that this can naturally be viewed as an optimal stopping problem with the objective of maximizing decoder efficiency. While aggressive aborting risks discarding recoverable executions, current designs introduce unnecessary latency and resource usage. Our aim is to optimize this trade-off.

To overcome the drawback of conventional FD-only design, we develop adaptive abort schemes that dynamically adjust how many syndrome measurement rounds to perform in each shot based on the syndromes observed so far (see Figure~\ref{fig:abort_shot}). If early rounds indicate that the shot is highly likely to result in an unrecoverable logical error, the system can abort the current shot and start a new shot. 
Aborting unpromising shots is crucial as measurement and decoding need significant time and hardware resources in a fault-tolerant run. By reducing these unpromising shots, the decoder can process less syndrome data and make fewer calls, allowing it to complete more useful decodes per unit time. 
In decoder-limited setups, this also reduces overall runtime by freeing qubit time and decoding cycles.
This approach is particularly advantageous because, on many quantum platforms, a fast physical reset of both ancilla and data qubits is substantially cheaper than completing a full syndrome measurement round~\cite{geher2025reset}.

To reduce the measurement and decoding latency, we introduce a fast, reliable decision module named \textit{Adaptive-Abort Module} that evaluates each shot in real time and aborts shots as they become more likely to fail (see Figure~\ref{fig:abort_module}).
The approach to building this module involves using syndrome measurement data to train Machine Learning (ML) and Reinforcement Learning (RL) models, which can then learn error patterns of quantum hardware. Using this approach, we introduced two adaptive abort schemes. First, an RL-based  \textit{One‑Step Lookahead (OSLA)} decoding scheme, at each round, compares the cost of stopping now against the cost of taking one more measurement round and then stopping. Second, an ML-based \textit{Adaptive-Threshold Abort (AdAbort)} decoding scheme estimates the logical-error probability at the time of decoding at each round. It terminates the shot as soon as that probability exceeds a threshold. 
By slotting this adaptive module between the decoder and the hardware interface, we can significantly reduce the processing at the decoder and the number of decoder calls by terminating unpromising shots early. Notably, the adaptive controller remains agnostic to the decoder and the code; these components remain unchanged. In this way, it can be integrated seamlessly into existing QEC architectures without modifying the underlying coding and decoding system.

Our contributions are as follows:
\begin{enumerate}
    \item Conceptually, we provide a new framing of quantum error correction as a sequential decision problem. Here, a controller must decide whether to continue or to adaptively abort the current execution. This provides a new perspective that differs from the traditional fixed-depth QEC approach.
    \item Methodologically, this study introduces two algorithms to implement the optimal stopping approach. The first utilizes a reinforcement learning method based on the one-step lookahead rule (OSLA), which continuously evaluates the cost of continuing versus stopping. The second, within the AdAbort framework, reformulates the problem as a machine learning inference task. The AdAbort framework involves learning the probability of decoding failure for a given instance and comparing this probability to a threshold to determine whether to stop or continue execution. These methods represent an initial step toward practical and implementable policies for adaptive execution in QEC environments.
    \item Empirically, we assess the FD, OSLA, and AdAbort schemes with Stim \cite{gidney2021stim} on surface codes and color codes with a circuit-level depolarizing noise model. OSLA and AdAbort deliver significant decoder efficiency gains over FD decoding. Importantly, as the code distance increases, efficiency increases. When compared with FD decoding, AdAbort reaches a relative improvement in efficiency of over $35\%$ (surface codes) and nearly $60 \%$ (color codes) at $d=15$.
    \item In practice, an abort module is a controller-level system that can be implemented without modifying the rest of the quantum computing stack. So our proposal can improve efficiency with minimal overhead. Further, the system is agnostic to the decoding and coding mechanisms used. It simply requires adaptive reaction to incoming syndrome information, and doing so can lead to a more effective system, particularly as the system size grows.
\end{enumerate}

Our results highlight that early abort decoding is a simple yet powerful module for any fault tolerant quantum computation. 



\subsection{Organization}
The rest of the paper is organized as follows. In Section~\ref{sec:related_works}, we discuss relevant works for our decoding problem. We discuss the background needed to understand the QEC code in Section~\ref{sec:background}, specifically surface and color codes. Furthermore, the concrete system model is described in Section~\ref{sec:sys_model}. We then discussed adaptive abort decoding schemes for QEC considered in this work in Section~\ref{sec:decoding_schemes}. Data generation and evaluation of the trained model are given in Section~\ref{sec:training_data}. For evaluating our work, the QEC benchmarks are discussed in Section \ref{sec:benchmark}. Evaluation results for surface codes and color codes are mentioned in Section~\ref{sec:numerical_results}. Finally, we conclude the paper in Section~\ref{sec:colculsion}.
\section{Related works}
\label{sec:related_works}       

Recently, decoding planar QEC codes has received significant attention within the QEC community, particularly for speed, accuracy, and real-time implementation. Several works have explored different aspects of decoding strategies for planar codes. The Minimum‐Weight Perfect‐Matching (MWPM) decoder remains a base for surface codes, offering a balance of speed and accuracy~\cite{kitaev2003fault,dennis2002topological,higgott2025sparse}. Alternative classical decoders such as the Union‑Find decoder~\cite{delfosse2021almost} and belief‐propagation approaches~\cite{roffe2005decoding} have also proven effective on topological codes. More recently, ML and RL methods have been applied to the decoding problem, yielding promising results for small to medium code distances~\cite{andreasson2019quantum,sweke2020reinforcement,fitzek2020deep,jung2024convolutional,baireuther2018machine,baireuther2019neural}. However, these ML/RL decoders become increasingly costly to train and deploy as the code distance grows. In contrast, our work focuses on a complementary question: rather than replacing the decoder itself, we introduce a lightweight abort module that sits alongside any existing decoder.
For more details on decoders for surface codes refer to ~\cite{demarti2024decoding}.

Several recent works~\cite{chamberland2023techniques,smith2023local,alavisamani2024promatch,ravi2023better,das2022lilliput} have explored speeding up QEC decoders by offloading or preprocessing the easy error patterns.
Promatch~\cite{alavisamani2024promatch} introduces a predecoder that prematches simple syndrome patterns on‐chip, reducing the burden on a full MWPM decoder and extending real‐time decoding to larger code distances.
Similarly, in~\cite{ravi2023better}, authors propose a two‐tier clique decoder that handles trivial cases entirely on‐chip, cutting off $70-99\%$ of off‐chip workloads without losing accuracy. Furthermore, in~\cite{das2022lilliput}, authors introduce a compact, on‑chip lookup‑table decoder "Lilliput" that handles low‐weight error patterns in a few clock cycles, and uses MWPM only for rare cases.
The lookup tables are prefilled offline by running a software decoder on every possible detection‐event pattern. 
These prior works focus on accelerating the full-depth decode; they reduce the cost of running MWPM (or avoid it entirely for common or easy cases) once the entire syndrome data of a shot is available. By contrast, our adaptive-abort approach saves work earlier in the pipeline. Rather than only making decoding faster after all rounds are measured, we use the incoming syndrome stream to predict whether a shot is unlikely to be recoverable and terminate the shot before completing the remaining measurement rounds or invoking a decoder. 

\section{Background on quantum error correction}
\label{sec:background}  
In this section, we review the background necessary to understand QEC in terms of both error detection and error correction. After describing basic concepts and terminology, we introduce surface and color codes, two of the standard coding schemes in QEC, after we discuss decoding. 

QEC codes protect logical information by repeatedly measuring a set of stabilizers commuting Pauli operators whose joint $+1$ eigenspace defines the code. In each syndrome measurement round, for every stabilizer an ancilla qubit is initialized (in $\lvert0\rangle$ for a $Z$-check or $\lvert+\rangle$ for an $X$-check), entangled with the data qubits via CNOT gates according to each stabilizer, and then measured.  The measurement outcome is the syndrome bit, which flags whether the measured stabilizer has experienced an odd parity of errors. After collecting all syndrome bits, a classical decoder processes the full syndrome pattern to infer the most likely pattern of bit‑flip and phase‑flip errors on the data qubits and outputs a set of corrective Pauli operators.  Applying these corrections, the code is restored to its logical state.  Finally, ancillas are reset in preparation for the next round.

\subsection{Terminology}
A QEC scheme consists of measurement rounds; a sequence of measurement rounds creates a shot, which is an attempt at a full circuit execution. In an adaptive scheme, a shot may or may not be aborted. Recall Figure \ref{fig:abort_shot}.

Specifically, a \textit{syndrome measurement round} includes all the operations necessary to measure each stabilizer exactly once.  In every round, ancilla qubits are first initialized, then entangled with the data qubits via CNOT gates associated with each stabilizer, and finally measured to produce syndrome bits that indicate where errors may have occurred.

A \textit{shot} is a single execution attempt for the entire QEC circuit from the first syndrome measurement round, through any subsequent rounds. The shot may be aborted after any round in an adaptive scheme, but if not, the shot will continue through all rounds, then to decoding, correction, and the final readout, yielding one complete set of measurement outcomes.

\subsection{QEC using surface code}
The rotated surface code~\cite{kitaev2003fault,fowler2012surface} is a topological QEC code defined on a two-dimensional lattice. It encodes a single logical qubit into $d^2$ data qubits for code distance $d$. The data qubits are arranged at the vertices of a rotated square grid, with ancilla qubits placed at the center of each plaquette for syndrome extraction. In total, there are $n = 2d^2-1$ physical qubits.

\begin{figure}[htb]        
  \centering
  \begin{subfigure}{0.48\columnwidth}
    \includegraphics[width=\linewidth]{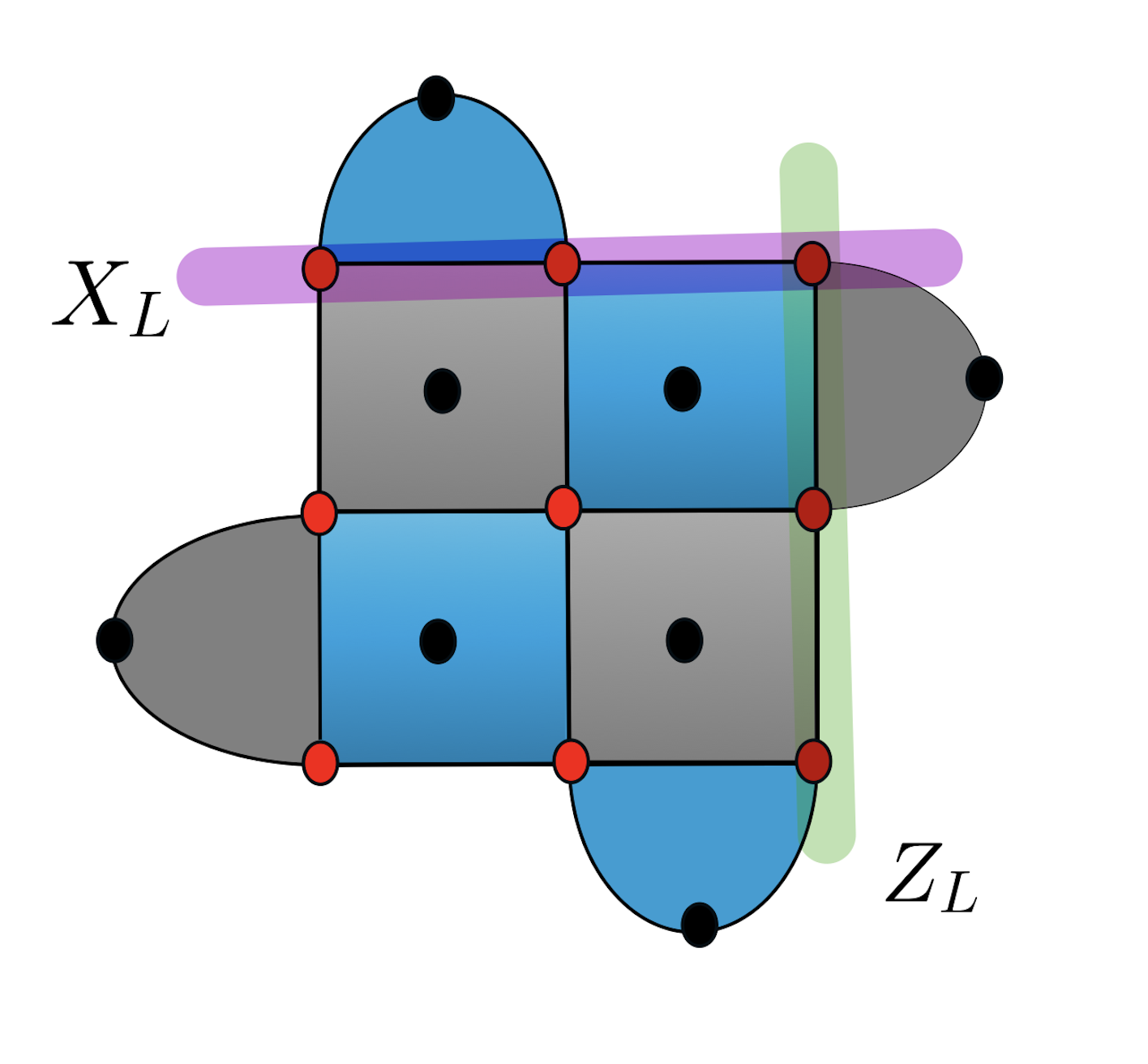}
    \caption{$[9,1,3]$ rotated surface code}
    \label{fig:roated_surface_code}
  \end{subfigure}\hfill
  \begin{subfigure}{0.48\columnwidth}
    \includegraphics[width=\linewidth]{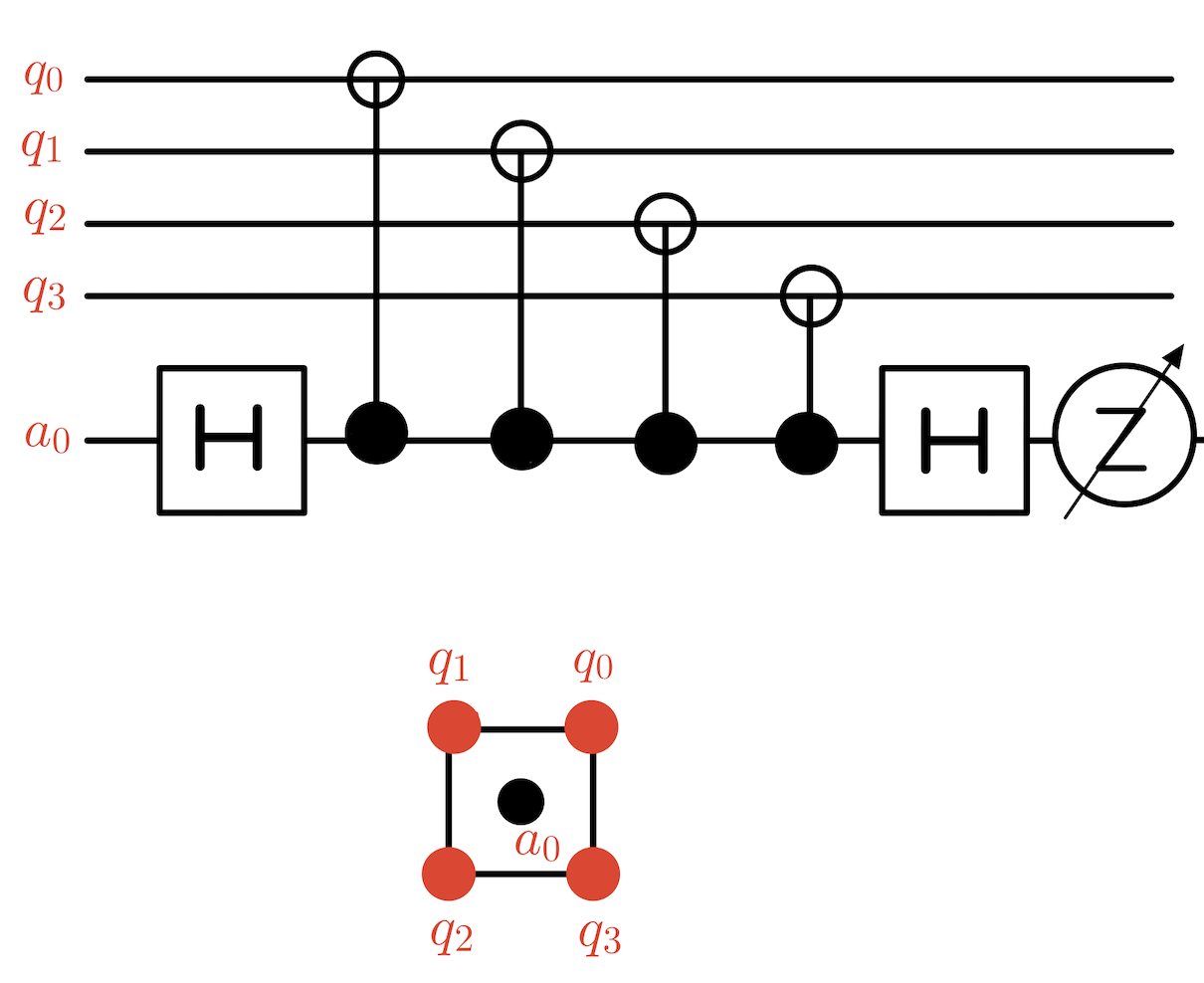}
    \caption{$X$-stabilizer measurement in rotated surface code}
    \label{fig:measuring_ckt}
  \end{subfigure}
  \caption{(a) Layout of the $[9,1,3]$ rotated‑surface code patch: red circles are data qubits, black circles are ancilla qubits; grey plaquettes indicate $X$‑type checks 
  and blue plaquettes indicate $Z$‑type checks. A pink chain of $X$ operators between the two \textit{rough} boundaries implements the logical $X_L$ operator, while a green chain of $Z$ operators between the \textit{smooth} boundaries implements the logical $Z_L$ operator. (b) Quantum circuit used to measure an $X$‑stabilizer in the rotated surface code.}
  \label{fig:side-by-side}
\end{figure}

In the rotated surface code, each plaquette is assigned either an $X$-type stabilizer or a $Z$-type stabilizer, with these two types alternating on adjacent faces. Figure~\ref{fig:roated_surface_code} illustrates the smallest nontrivial instance, the $[9,1,3]$ rotated surface code, where red circles are data qubits, black circles are ancillas for stabilizer measurement, blue plaquettes are $Z$-stabilizers, and grey plaquettes are $X$-stabilizers. To measure an $X$-stabilizer on a given plaquette (grey colour in Figure~\ref{fig:roated_surface_code}), we perform an ancilla‑controlled parity check of the four data qubits around that face, effectively applying the product of their Pauli‑$X$ operators (see Figure~\ref{fig:measuring_ckt}). Similarly, a $Z$-stabilizer (blue colour
 in Figure~\ref{fig:roated_surface_code}) measures the combined parity of their Pauli‑$Z$ operators.
 Because every $X$-type stabilizer overlaps each $Z$-type on an even number of qubits, all stabilizers commute and jointly define the code space. 
 By repeating these measurements across all plaquettes, the code converts any local error into a distinctive syndrome pattern. A classical decoder reads this syndrome pattern to pinpoint and correct the error without ever collapsing the logical quantum state.

 In a distance‑$d$ rotated surface code, the code can detect any error pattern of fewer than $d$ physical errors. However, once $d$ or more errors occur along a connected path from one boundary to the opposite, they together form an uncorrectable chain and induce a logical error on the encoded qubit.
These logical errors take the form of continuous strings of Pauli $X$ or $Z$ operators running between opposite edges. An $X$-string connecting the two rough boundaries shown in pink in Figure~\ref{fig:roated_surface_code} implements the logical $X_L$ flip. Similarly, a $Z$-string connecting the smooth boundaries shown in green implements the logical $Z_L$ flip.

\subsection{QEC using color codes}
\begin{figure}[h!]
\centering
  \medskip
  \centering
  \includegraphics[width=5cm]{ 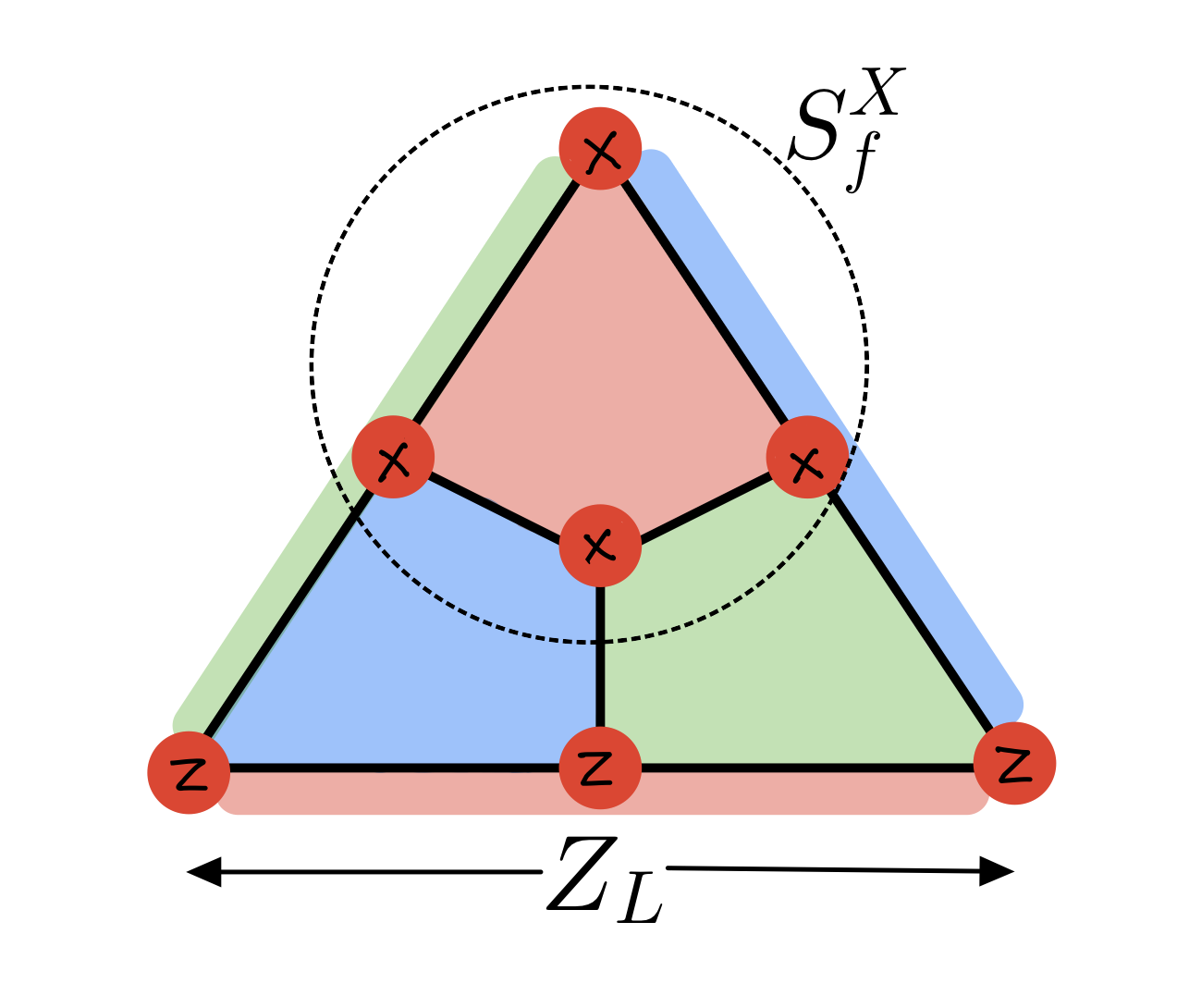}
\caption{The $[7,1,3]$ triangular color code. Data qubits (red circles) reside on the vertices, with each of the three triangular faces shaded with red, green, and blue colors. An example $X$-type stabilizer $S_f^X$ on the top (red) face. The logical $Z$ operator $Z_L$ is realized by the product of $ Z$'s on the three qubits along any one colored boundary in this illustration, the bottom red edge.
}
\label{fig:colorcode}
\end{figure}

The quantum color code~\cite{bombin2006topological,fowler2011two,landahl2011fault} is also a planar topological QEC code defined on trivalent, $3$‑colorable lattices in which every vertex has degree three and faces are colored in three colors (red, green, and blue) so that no two adjacent faces share the same color (see Figure~\ref{fig:colorcode}). Data qubits reside on the vertices of the lattice, and end each face $f$ gives rise to two Pauli‐check stabilizers $S_f^X$ for $X$-type stabilizer and $S_f^Z$ for $Z$-type stabilizer acting on the set of boundary data qubits of face $f$. 
In practice, color codes are often realized using two types of planar patches, namely triangular patches and hexagonal patches. A triangular patch consists of a compact triangular arrangement of hexagonal faces with three distinct colored boundaries, and it encodes one logical qubit at distance $d$. 

The logical operator $X_L$  and logical operator $Z_L$ in a triangular color code are realized as a string of single qubit Paulis acting on all data qubits along one of the three colored boundaries. As shown in Figure~\ref{fig:colorcode}, the logical $Z_L$ operator is realized by a connected string of Pauli
$Z$ operators acting on every data qubit along red-colored boundary of the triangular patch.

An important property of color codes is that the entire Clifford group is implemented transversally, that is a logical Hadamard is simply bitwise $H$, a logical phase $S$ is bitwise $S$, and logical CNOT is a layer of physical CNOTs between matching qubits of two code blocks.

\subsection{Decoding error pattern}
Once we obtain the syndrome bits from the stabilizer measurements, classical decoding algorithms can be used to predict the most likely error pattern on the data qubits. The commonly used decoder for the planar codes is the MWPM~\cite{kitaev2003fault}.  Errors on data qubits light up a pair of syndrome measurement bits on adjacent check locations. Using these bits, one can construct a graph whose vertices are the observed defect locations and edges connect every pair of defects, weighted by the shortest path length (Manhattan distance) between them on the dual lattice. The decoder then finds the matching of these defects in a way such that no error vertex is left unconnected and the sum of the edge weights is minimized using Edmonds' blossom algorithm~\cite{edmonds1965paths}. The matched edges then define the error chain, and when applied as corrections, will most likely revert the state of the logical qubit to the code space without leading to any logical error.



\section{System model}
\label{sec:sys_model}
Let $T_d$ denote the number of syndrome measurement rounds in a shot for code distance~$d$. In our model, we perform syndrome measurements in discrete rounds labeled $t = 1,2,\dots,T_d$. At each round $t>0$, we collect a syndrome bit vector
$$
\bm s_t \in \{0,1\}^{n_{\rm checks}},
$$
where $n_{\rm checks}$ is the total number of stabilizer generator measurements (for a distance $d$ rotated surface code, $n_{\rm checks} = d^2 - 1$). We assume that all $n_{\rm checks}$ stabilizers are measured in parallel during round $t$, so the time it takes to complete a round does not grow with $n_{\rm checks}$. We write the syndrome measurement history up round $t$ as
$$
\bm X_t = \bigl(\bm s_1,\,\bm s_2,\,\dots,\,\bm s_t\bigr).
$$
If no early abort occurs, we continue until $t = T_d$, then pass the full history $\bm X_{T_d}$ to a classical decoder $\mathcal{D}$ that infers the most likely pattern of bit‑flip and phase‑flip errors on the data qubits. The decoding schemes introduced in this work are applied before any classical decoder $\mathcal{D}$ and are therefore agnostic to the choice of decoder. 


Each syndrome measurement round takes a fixed amount of execution time $M$ in proper units. Specifically, $M$ includes execution time of one layer of parallel CNOTs, ancilla readout, and fast reset of all ancilla qubits.  We assume that there is a constant measurement time per round, independent of the number of stabilizers.
If, at some round $t<T_d$, we decide to abort the shot rather than continue to round $t+1$, we pay an additional reset time cost $R_{\mathrm{reset}}>0$ for aborting the shot and reinitializing the logical qubit.
In practice, $R_{\mathrm{reset}}$ is smaller than $M$, since resetting the logical qubit and reinitializing ancillas is cheaper than running a whole extra round (See discussion in the Introduction).  Thus, if we abort the shot at $t=\tau $, the total time spent on that shot is $\tau M + R_{\mathrm{reset}}$.

We also include a nonzero decoder failure cost $D_{\text{fail}}$ because shots that reach the decoder but still fail impose an overhead beyond simple measurement and reset times.  In practice, a logical failure triggers extra classical processing to detect and diagnose the error, and delays for re‐initializing the code lattice.  By charging each failed attempt by the decoder an additional cost $D_{\rm fail}$, our model reflects the penalty incurred by unsuccessful decoding, rather than treating all failures as costless.


 \begin{rem}
In adaptive abort schemes, when a shot is stopped, it is immediately aborted, and its syndromes are discarded. Therefore, no decoding or recovery is applied. Only shots that continue until the full depth $T_d$ of syndrome rounds are passed to the decoder and counted toward the decoder efficiency, formally defined later.
 \end{rem}

\section{Adaptive abort schemes}
\label{sec:decoding_schemes}

To reduce the latency of the decoder, we dynamically adjust the number of syndrome measurement rounds per shot instead of always using a fixed $T_d$.  If early syndrome data indicate that the current shot is likely to produce a logical error with high probability, the system can abort it immediately. By terminating a failing shot, resetting the logical qubit, and starting a new execution shot, we can avoid spending time on additional rounds within the aborted shot. In this work, we proposed two adaptive decoding schemes and compared their performance with the existing FD scheme. 


\subsection{Fixed-depth (FD) decoding }

The FD decoding scheme is the standard approach used in current QEC controller architectures (see Figure~\ref{fig:fd_module}) and has been the baseline in both theoretical and experimental studies~\cite{terhal2015quantum,fowler2012surface,kitaev2003fault,barends2014superconducting}.
The FD scheme chooses a maximum number of rounds, $T_d$, and always performs exactly $T_d$ syndrome measurement rounds before applying the decoder. 
Specifically, it collects full syndrome measurement history $\bm X_{T_d}$. Only after completing all $T_d$ rounds, the complete syndrome history $\bm X_{T_d}$ is sent to the decoder $\mathcal{D}$. 
If $\mathcal{D}(\bm X_{T_d})$ indicates no logical error, the shot is declared successful; otherwise, it is a failure. 
Since there is no early‐abort option, the execution cost per shot is always $T_dM$.

\subsection{One-step look ahead (OSLA) decoding}
The first adaptive decoding scheme considered in this work is the Reinforcement Learning (RL) based One-Step Lookahead (OSLA) decoding. Under the OSLA decoding scheme, at each round, it decides whether to stop immediately or to take one more syndrome round before deciding. To make that choice, it estimates two probabilities based on the current syndrome measurement round $\bm s_t$: (a) the current logical error probability estimate $g(\bm s_t)$, which is the probability of a logical error if stop now (b) the expected future logical error probability estimate $m(\bm s_t)$, which is expected $g(\bm s_{t+1})$ after doing one more round. Formally, the OSLA decoding scheme uses estimates of logical error probabilities

\begin{align}
&g(\bm s_t) = \Pr(\text{logical error if we stop at }t \mid \bm s_t),\label{eqn:l_1}\\
&m(\bm s_t) = \mathbb{E}[g(\bm s_{t+1}) \mid \bm s_t].\label{eqn:l_2}    
\end{align}
We consider a continuation cost $c(\bm{s}_t)<0$, which reflects the fact that performing another round can decrease the expected logical error probability.  
For a running shot, if at round $t$ we have 
$$
c\bigl(\bm{s}_t\bigr)+m\bigl(\bm{s}_t\bigr)>g\bigl(\bm{s}_t\bigr),
$$
then it is better to stop now and incur the execution cost of $tM + R_{\mathrm{reset}}$ for the measurement rounds already used and resetting the logical qubit. Otherwise, continue to round $t+1$.  If no early stop occurs by $t = T_d$, pay $T_d\,M$ for all measurements and then decode $\mathcal{D}(\bm{X}_{T_d})$.  In this way, the OSLA decoding scheme automatically balances the continuation benefit $c(\bm{s}_t)$ against the expected reduction in logical‐error probability $m(\bm{s}_t)-g(\bm{s}_t)$. In several settings, OSLA is known to be optimal. The pseudo code for the OSLA decoding scheme is given in Algorithm~\ref{alg:osla}. 

Note that OSLA is an RL policy: at each syndrome round, it must choose between stopping and measuring one more round.  It relies on learned estimates $g(\bm X_t)$ and $m(\bm X_t)$ to perform a one‑step Bellman update, selecting the action with the lower expected total cost.  In other words, OSLA is a greedy, value‑based RL method that minimizes latency by looking just one round ahead. For more details on optimal stopping and reinforcement learning, see~\cite{bertsekas2012dynamic,tsitsiklis1996approximate}. 

\begin{algorithm}[h]
\caption{OSLA decoding}
\label{alg:osla}
\begin{algorithmic}[1]
\REQUIRE $T_d$ rounds, trained model for $g(\cdot)$ and $m(\cdot)$, $c<0$
\REQUIRE subroutines: \texttt{measure()}, \texttt{reset()}, \texttt{decode($\bm X_{T_d}$)}
\FOR{shot = 1 to $N$}
  \STATE $\texttt{aborted} \leftarrow \text{false}$
  \FOR{$t = 1$ to $d-1$}
    \STATE $\bm s_t \leftarrow \texttt{measure()}$
    \STATE append $\bm s_t$ to make $\bm X_t$
    \STATE $g_t \leftarrow g(\bm s_t)$
    \STATE $m_t \leftarrow m(\bm s_t)$
    \IF{$c + m_t > g_t$}
      \STATE $\texttt{aborted} \leftarrow \text{true}$
      \STATE \texttt{reset()} \COMMENT{reset and next shot}
      \STATE \textbf{break}
    \ENDIF
  \ENDFOR
  \IF{not \texttt{aborted}}
    \STATE $\bm s_d \leftarrow \texttt{measure()}$
    \STATE append $\bm s_d$ to make $\bm X_{T_d}$
    \STATE $\texttt{result} \leftarrow \texttt{decode}(\bm X_{T_d})$
    \STATE \texttt{record\_outcome(result)}
  \ENDIF
\ENDFOR
\end{algorithmic}
\end{algorithm}

\subsubsection{Model to predict $g(\cdot)$ and $m(\cdot)$ for OSLA decoding}

To train the OSLA decoding scheme with both “stop” and “continue” actions, we simulate syndrome histories 
under the noise model discussed in~\ref{sec:error_model}. This directly encodes both decision scenarios, stop now versus measure one more, so the model can compare their expected total costs at each step.
Since neural networks require fixed-size inputs, each trajectory of length $1$ is padded to the maximum length ($2$ rounds) using an appropriate padded value. We then construct two label targets, both set equal to the true logical-error outcome observed at the end of the syndrome trajectory. The neural network used for OSLA (Table~\ref{tab:two_head_arch_compressed}) has two sigmoid outputs:
\begin{itemize}
    \item \textit{Stop-now sigmoid output (predicted $g(\cdot)$)}: Estimates the probability of logical failure if stopped immediately.
    \item \textit{Next-step sigmoid output (predicted $m(\cdot
    )$)}: Estimates the probability of logical failure after one additional round of syndrome measurement.
   Both heads are trained jointly, so that the network learns the cost estimates required for OSLA’s one-step lookahead decision at each time step.
\end{itemize}

\begin{table}[h!]
  \centering
  \scriptsize
  \resizebox{0.8\columnwidth}{!}{%
    \begin{tabular}{lcc}
      \toprule
      \textbf{Layer Type}      & \textbf{Output Size}            & \textbf{\# Parameters} \\
      \midrule
      Input                    & $(T_d,n_{\rm checks})$  
                               & —  \\
      Flatten                  & $(T_d n_{\rm checks},)$ 
                               & —  \\
      Dense + ReLU             & $(128,)$                         
                               & $128(T_dn_{\rm checks}) + 128$ \\
      Dense + ReLU             & $(64,)$                          
                               & $8256$ \\
      \midrule
      \textit{g (sigmoid)}& $(1,)$                           
                               & $65$ \\
      \textit{m (sigmoid)}& $(1,)$                           
                               & $ 65$ \\
      \bottomrule
    \end{tabular}%
  }
  \caption{Architecture of the OSLA model} 
  \label{tab:two_head_arch_compressed}
\end{table}

\subsection{AdAbort decoding}
Note that the OSLA decoding scheme makes decisions based on the cost trade-off between stopping now and taking exactly one more measurement round. This makes OSLA very lightweight, requiring only simpler training data and a simple model. However, it is also inherently short-sighted, as it cannot anticipate what might happen after two or more rounds.
Therefore, the second adaptive decoding scheme we consider in this work is the ML‑based AdAbort for the abort module shown in Figure~\ref{fig:abort_module}. The AdAbort decoding scheme approximates the conditional probability of logical error at the final round $T_d$ using a neural‑network function approximator that takes the current syndrome history $\bm X_t$ as input and outputs an estimated error probability. We denote this by

\begin{equation}
\hat{p}_{\mathrm{err}}(\bm X_t) = \Pr\bigl(\text{logical error at round }T_d \,\bigm|\, \bm X_t\bigr).   
\end{equation}
At each syndrome measurement round $t \le T_d$, the AdAbort scheme first computes $\hat{p}_{\mathrm{err}}(\bm X_t)$ and compares it to a threshold $\theta\in (0,1)$. If
$
\hat{p}_{\mathrm{err}}(\bm X_t) \ge \theta,
$ the current running shot is aborted. In that case, having completed $t$ rounds (with execution time $tM$) and paying the reset cost $R_{\mathrm{reset}}$, the total execution time for the aborted shot is $t M + R_{\mathrm{reset}}$. After aborting the shot, the logical qubit is reset, and the execution of a new shot begins. On the other hand, if $p_{\mathrm{err}}(\bm X_t) < \theta$, we proceed to the next round. If no abort occurs before $t = T_d$, then all $T_d$ rounds are completed with total execution time of $T_dM$ and the decoder $\mathcal{D}(\bm X_{T_d})$ is used for decoding. The pseudo code for the AdAbort decoding scheme is given in Algorithm~\ref{alg:adabort}. 
An aborted shot costs $t\,M + R_{\mathrm{reset}}$, while a shot that runs all $T_d$ rounds costs $T_dM$ before decoding.
 \begin{rem}
     The threshold $\theta$ balances speed and fidelity: lowering $\theta$ aborts more shots early (faster but fewer completed shots), while increasing $\theta$ preserves more shots to full decode (higher fidelity but longer latency). By calibrating $\theta$, one can guarantee the logical‐error rate stays below any target. For example, setting $\theta=0.05$ means any shot whose predicted failure probability exceeds $5\%$ is aborted, ensures that at most $5\%$ of completed shots suffer a logical error (at least $95\%$ fidelity), while still aborting many hopeless shots early to save latency.
 \end{rem}


\begin{algorithm}[h!]
\caption{AdAbort decoding}
\label{alg:adabort}
\begin{algorithmic}[1]
\REQUIRE $T_d$ rounds, trained model for $\hat{p}_{\mathrm{err}}(\cdot)$, threshold $\theta$
\REQUIRE subroutines: \texttt{measure()}, \texttt{reset()}, \texttt{decode($\bm X_{T_d}$)}
\FOR{shot = 1 to $N$}
  \STATE $\texttt{aborted} \leftarrow \text{false}$
  \FOR{$t = 1$ to $T_d$}
    \STATE $\bm s_t \leftarrow \texttt{measure()}$ \COMMENT{Perform syndrome round}
    \STATE append $\bm s_t$ to make $\bm X_{t}$
    \STATE $\hat{p}_{\mathrm{err}}(\bm X_t) \leftarrow \texttt{model.predict}(\bm X_{t})$
    \IF{$\hat{p}_{\mathrm{err}}(\bm X_t) \ge \theta$}
      \STATE $\texttt{aborted} \leftarrow \text{true}$
      \STATE \texttt{reset()} \COMMENT{reset and next shot}
      \STATE \textbf{break}
    \ENDIF
  \ENDFOR
  \IF{not \texttt{aborted}}
    \STATE $\texttt{result} \leftarrow \texttt{decode}(\bm X_{T_d})$
    \STATE \texttt{record\_outcome(result)}
  \ENDIF
\ENDFOR
\end{algorithmic}
\end{algorithm}

\subsubsection{Model to predict $\hat{p}_{\mathrm{err}}(\cdot)$ for AdAbort decoding}
To train our AdAbort decoding scheme, we do the following three steps: (1) generating noisy syndrome sequences, (2) augmenting these sequences by extracting all prefixes, and (3) training a convolutional neural network (CNN) to predict the probability of logical failure from any partial syndrome history.

An important feature of AdAbort decoding is that at each round $t$ it decides, using only the first $t$ syndrome rounds, whether to abort or continue. To train a model that can operate for every $t\in\{1,\dots,T_d\}$, we expand each full $T_d$-round trajectory into $T_d$ training examples. For shot $i$, let $\bm{S}_i\in\{0,1\}^{T_d\times(d^2-1)}$ denote the syndrome matrix whose $r$-th row $\bm{s}_i^{(r)}\in\{0,1\}^{d^2-1}$ is the syndrome bit-vector measured at round $r$, and let $y_i$ be its label. For each decision time $t$, we form a fixed-shape input $\tilde{\bm{S}}_i^{(t)}\in\mathcal{A}^{T_d\times(d^2-1)}$ by keeping the first $t$ rows and padding the remaining $T_d-t$ rows with a designated padding value $s_{\mathrm{pad}}\in\mathcal{A}$:
\[
\tilde{\bm{S}}_i^{(t)}[r,j] =
\begin{cases}
\bm{S}_i[r,j], & 1\le r\le t,\\
s_{\mathrm{pad}}, & t<r\le T_d,
\end{cases}
\qquad
r=1,\dots,T_d,\ \ j=1,\dots,d^2-1,
\]
where $\mathcal{A}\triangleq\{0,1\}\cup\{s_{\mathrm{pad}}\}$ denotes the augmented alphabet including the padding symbol. The resulting training set is $\{(\tilde{\bm{S}}_i^{(t)},y_i)\}_{t=1}^{T_d}$. The model used to train the AdAbort decoding scheme is described in Table~\ref{tab:cnn_arch_numeric}.

\begin{table}[h!]
  \centering
  \caption{CNN architecture to train AdAbort scheme}
  \label{tab:cnn_arch_numeric}
  \begin{tabular}{lcc}
    \toprule
    \textbf{Layer Type}                & \textbf{Output Size}         & \textbf{\# Parameters} \\
    \midrule
    Input                              & \((T_d,\,n_{\rm checks})\)     & —        \\
    Permute                            & \((n_{\rm checks},\,T_d)\)     & —        \\
    Conv1D + ReLU (64,3)    & \((n_{\rm checks},\,T_d)\)     & 12352   \\
    BatchNormalization                 & \((n_{\rm checks},\,T_d)\)     & 128      \\
    Conv1D + ReLU (64,3)    & \((n_{\rm checks},\,T_d)\)     & 12352   \\
    BatchNormalization                 & \((n_{\rm checks},\,T_d)\)     & 128      \\
    Permute                            & \((T_d,\,64)\)                 & —        \\
    GlobalAveragePooling1D             & \((64,)\)                    & —        \\
    Dense + Sigmoid                    & \((1,)\)                     & 65       \\
    \bottomrule
  \end{tabular}
\end{table}

\section{Training data generation, labelling, and model evaluation}
\label{sec:training_data}


For each code distance $d$, we generate $T_d$ rounds of syndrome extraction per shot using \textsc{Stim}, where we set $T_d=d$ following the standard choice in syndrome-based decoding: repeating stabilizer measurements for $d$ rounds provides a space-time volume of sufficient depth to reliably identify and correct faults, consistent with the distance-$d$ fault-tolerance criterion for phenomenological and circuit-level noise models. We then simulate a circuit-level depolarizing noise model. Specifically, immediately before each syndrome-measurement round, each data qubit is independently subjected to a single-qubit depolarizing channel with probability $p$. After each entangling \textsc{CNOT} gate used for stabilizer measurement, we apply an independent two-qubit depolarizing channel with probability $p$.

\subsection{Definition of logical error and label construction}
A logical error is defined as a failure event in which the decoded logical observable at the end of the memory experiment differs from the prepared logical state. For example, in a memory-$X$ experiment (where the logical $|+\rangle_L$ state is prepared and the logical $X$ operator is measured at the end), a logical error occurs if the final measured logical $X$ eigenvalue is flipped relative to the ideal outcome.
In our implementation to match the downstream decoding pipeline, we additionally compute whether the chosen classical decoder (MWPM for surface codes) correctly predicts the logical observable from the measured detector events. Concretely, for each shot $i$, we run a classical decoder on the detector outcomes to obtain a predicted logical-observable value $\hat{o}_i$, and compare it to the ground-truth observable $o_i$ produced by the simulator. We then assign a binary label
\begin{equation}
\label{eqn:label}
y_i =
\begin{cases}
1 & \text{if } \hat{o}_i = o_i \quad(\text{decoder succeeds}),\\
0 & \text{if } \hat{o}_i \neq o_i \quad(\text{decoder failure}).
\end{cases}   
\end{equation}
This makes the training target directly reflect whether full-depth decoding would succeed, which is precisely what the abort module is designed to anticipate.

\subsection{Data generation for AdAbort decoding} For a fixed code distance $d$ and number of rounds $T_d$, each Monte-Carlo shot $i$ yields a binary syndrome history
$
\bm S_i \in \{0,1\}^{T_d \times n_{\rm checks}},
$
where the second dimension is the number of check outcomes per round (for a rotated surface-code $n_{\rm checks}=d^2-1$).  
Each shot also yields a single binary final outcome label $y_i \in \{0,1\}$ indicating whether the shot ended in a logical failure under the chosen definition~\eqref{eqn:label}. 


AdAbort requires predictions after any intermediate round $t$, so we convert each full shot into $T_d$ supervised examples by using all syndrome prefixes and padding them to a fixed shape. Specifically, for shot $i$ with syndrome matrix $\bm S_i\in\{0,1\}^{T_d\times n_{\rm checks}}$ and final label $y_i$, we construct, for each $t=1,\dots,T_d$,
\[
\bm X_{i,t} \;=\; \mathrm{pad}\!\left(\bm S_i[1{:}t]\right)\in \mathcal{A}^{T_d\times n_{\rm checks}},
\]
where $\bm S_i[1{:}t]\in\{0,1\}^{t\times n_{\rm checks}}$ denotes the submatrix consisting of the first $t$ syndrome rounds (i.e., the first $t$ rows of $\bm S_i$), and $\mathrm{pad}(\cdot)$ appends rows filled with a designated padding value $s_{\mathrm{pad}}$ in rounds $(t+1,\dots,T_d)$ so that the input shape is constant across all $t$. Here $\mathcal{A}\triangleq\{0,1\}\cup\{s_{\mathrm{pad}}\}$ is the augmented alphabet including the padding symbol. We assign the same final label to every prefix:
\[
\mathrm{label}(\bm X_{i,t}) \;=\; y_i.
\]
This produces $N\times T_d$ training examples from $N$ simulated shots.


For each $(d,p)$, we simulate $N$ independent shots to obtain syndrome label pairs $\{(\bm S_i,y_i)\}_{i=1}^{N}$. We split the shots into training and validation sets, expand each shot into padded prefixes $\{\bm X_{i,t}\}_{t=1}^{T_d}$ as described above, and train a classifier to output a probability estimate for the corresponding label. Performance is reported on the validation set using the ROC--AUC metric.

\subsection{Data generation for OSLA decoding}
For each physical error rate $p$ and code distance $d$, OSLA is trained on short, variable-length syndrome trajectories of length $1$ or $2$ rounds. The training set is constructed as follows.

For each training shot $i \in \{1,\dots,N\}$, we draw an integer number of rounds
$
r_i \sim \text{Uniform}\{r_{\min}, r_{\max}\},
$
with $r_{\min}=1$ and $r_{\max}=2$. This produces a mixed dataset containing both one-round and two-round shots. This construction is required because OSLA is a one-step lookahead policy: at a given time $t$ it must decide whether to stop now or take one more syndrome-measurement round by comparing the estimated logical-failure probability using the currently available syndrome information $\bm s_t$ (using~\eqref{eqn:l_1}) against the estimated logical-failure probability after acquiring the next round $\bm s_{t+1}$ (using~\eqref{eqn:l_2}). Therefore, the training set must contain both single-round and two-round shots so that the model can learn a stop-now estimate from $t$ rounds and a one-more estimate from $t+1$ rounds under the same noise conditions.

We then simulate a distance $d$ memory experiment for exactly $r_i$ rounds and sample one shot using Stim’s detector sampler. This returns a flattened binary detector-event vector of length $r_i n_{\rm checks}$ and a single observable-flip bit $y_i\in\{0,1\}$ used as the training label. The detector event vector is reshaped into a round-by-round syndrome vector $\bm S_i\in\{0,1\}^{r_i\times n_{\rm checks}}$, and to obtain a fixed input shape for the neural network, we post-pad $\bm S_i$ to $(r_{\max}, n_{\rm checks})$ using a appropriate pad value that lies outside the valid syndrome alphabet $\{0,1\}$.    

\subsection{Training Cost and Scalability}
   The training cost of both AdAbort and OSLA is dominated by data generation rather than neural-network size, and it scales polynomially with code distance. For AdAbort at distance $d$ for surface code, each shot produces a syndrome history of shape $T_d \times (d^2-1)$ with $T_d=d$; augmentation then converts each shot into $T_d$ supervised examples, so the total number of training examples scales as $O(Nd)$ for $N$ shots, while the syndrome per shot scales as $O(d^3)$. The CNN used for AdAbort is lightweight (fixed filter sizes and a small number of layers), so its parameter count does not grow exponentially with $d$; runtime increases mainly because larger $d$ implies more checks per round and more rounds per shots. In contrast, OSLA limits shots to one or two rounds. The two-head MLP is also small, and training typically converges quickly because the dataset is simpler and the sequence length is fixed at two. In practice, for both schemes the incremental cost of moving to larger code distances is driven by the increased number of stabilizer outcomes per round, rather than by an exploding model complexity, supporting scalability of the approach provided sufficient shots are generated for each $(d,p)$.
 
\subsection{Performance of the trained models at fixed $p=0.001$}
\label{sec:nn_performance_p0001}

We evaluate the trained neural predictors for AdAbort and OSLA at a fixed physical error rate $p=0.001$ across multiple code distances $d$ for the rotated surface code. For each code distance, we compute the ROC curve on a validation set and report the corresponding area under the ROC curve (ROC-AUC). 

For AdAbort, ROC-AUC quantifies how well the network separates shots that ultimately lead to logical failure from those that do not, using syndrome-history as inputs. For OSLA, we report ROC-AUC for both output heads: the stop-now head and the one-more head. Table~\ref{tab:roc_p0001} summarizes the ROC-AUC values. ROC-AUC values substantially above $0.5$ indicate that partial syndrome information contains predictive signal about eventual logical failure; comparing ROC-AUC across $d$ demonstrates how the learned predictors scale with code distance.

\begin{table}[t]
\centering
\caption{ROC-AUC at fixed $p=0.001$ for rotated surface code of multiple code distances.}
\label{tab:roc_p0001}
\begin{tabular}{c c c c}
\hline
 $d$ & AdAbort & OSLA $g(.)$ & OSLA $m(.)$\\
\hline
5  & 0.91 & 0.86 & 0.83 \\
7  & 0.89 & 0.84 & 0.81 \\
9  & 0.87 & 0.82 & 0.79 \\
\hline
\end{tabular}
\end{table}


\section{QEC benchmarks} 
\label{sec:benchmark}
In this section, we discuss the different benchmarking methods that we will use to evaluate our decoding schemes.
\subsection{QEC Codes}
To benchmark our decoding strategies, we simulate two families of topological codes: rotated surface codes and triangular-patch color codes.  For each code family, we considered the code distance $d\in\{3,5,7,9,11,15\}$. Surface‐code simulations employ a standard MWPM decoder, while color code benchmarks use a concatenated MWPM decoder optimized for triangular patches~\cite{lee2025color}.

For each code and each physical error rate $p$, we use \texttt{Stim}~\cite{gidney2021stim} to build a distance $d$ syndrome extraction circuit that applies single-qubit depolarizing noise before each round and two-qubit depolarizing noise after each Clifford gate (both with probability $p$).  We then perform $T_d=d$ rounds of stabilizer measurements per shot and repeat this process for $N$ independent shots to gather syndrome histories and logical‐flip outcomes. These datasets are then used in FD, OSLA, and AdAbort evaluation pipelines under the noise model detailed in Subsection~\ref{sec:error_model}.

\subsection{Performance metric}
To compare the three decoding schemes, we use a decoder efficiency, defined as the fraction of correct decoder outputs among shots that reach the decoder, divided by the average time per shot. Suppose we run $N$ shots for a fault-tolerant task, out of which $ N_s$ shots are successfully decoded by $\mathcal{D}$, $N_f$ shots reach decoding but fail, and $N_a$ shots are aborted early (do not reach the decoder). We define the success rate conditional on completion as $S_{\mathrm{rate}} = N_{s}/(N_{s} + N_{f})$, which is the probability that any shot reaching the decoder is correctly decoded. 
At the same time, each shot, whether it is aborted early or run to full depth, uses measurement time for its syndrome rounds and, if aborted, additional reset time. Denote the total execution time, including measurement time, reset time and decoder failure time for all $N$ shots by $T(N)$ and its expected value by $\mathbb{E}[T(N)]$.  We then define decoder efficiency ($\eta_\mathrm{dec}$) as the fraction of correctly decoded shots divided by the average time per attempted shot:

\begin{equation}    
\label{eqn:thpt}
\eta_\mathrm{dec} = \frac{S_{\mathrm{rate}}} {(\mathbb{E}[T(N)]/N)}.  
\end{equation}
The quantity $\eta_\mathrm{dec}$ captures the trade-off between correctness and resource use: 
it rewards schemes that both maintain a high conditional success probability and keep the average per-shot cost low. Rather than optimizing raw logical fidelity, we aim to optimize logical decoding throughput.
The quantity $\eta_\mathrm{dec}$ captures the trade-off between correctness and resource use: it rewards schemes that both maintain a high conditional success probability and keep the average per-shot cost low. From a decoding perspective, we care about the probability that a set of syndrome measurements entered into the decoder will yield a successful shot. For this reason, aborted shots are not included in the conditional success probability, as they are not fed into the decoder. Of course, there is some price in terms of aborting shots. This is accounted for in the time metric, where we must still pay a price in time to reset the system when aborting a shot. Thus, both factors are accounted for in the above efficiency metric.

Further, this is important from the decoder overhead perspective as adaptive schemes reduce the number of decoder calls by aborting error-prone shots, which directly lowers memory pressure and queueing delays in processing shots. Fewer decode calls reduce contention for limited decoder instances and lower the requirement for costly parallel hardware.


Note that for the FD scheme, the total execution time $T(N)$ for $N$ shots is deterministic because every shot always runs the same number of measurement rounds. By contrast, OSLA and AdAbort introduce stochastic stopping: some shots are aborted early while others run to full depth, so $T(N)$ is a random variable and we must use its expectation in~\eqref{eqn:thpt}. 

\subsection{Superconducting‐qubit timing parameters}
In our simulations, we considered a cost model for superconducting qubit hardware: each syndrome‐measurement round incurs $M = 0.7\mu\text{s}$ of readout, every aborted shot pays an additional reset overhead $R_{\rm reset} = 0.5\mu\text{s}$, and any decoding failure adds a classical processing latency $D_{\rm fail} = 1\mu\text{s}$. We simulate a large batch of shots and record the decoder efficiency $
  \eta_\mathrm{dec}
$.
Because the only hardware‐specific inputs are the three timing parameters $(M, R_{\rm reset}, D_{\rm fail})$, this entire framework can be immediately applied to other platforms such as trapped ions.


\subsection{Error Model}
\label{sec:error_model}
We model errors at the circuit level using two principal depolarizing channels, capturing both qubit idling and gate imperfections.  Let $d$ be the code distance and $T_d$ the total number of syndrome measurement rounds.  


Before the first CNOT of round $t$, each data qubit undergoes an independent single‐qubit depolarizing error with probability $p$.  In density‐matrix form, a data‐qubit state $\rho$ is replaced by
$
     (1 - p)\rho + \frac{p}{3}\bigl( X\rho X +  Y\rho Y +  Z\rho Z\bigr).
   $
Here, $X$, $Y$, and $Z$ denote the standard Pauli operators, where $X$ implements a bit‑flip, $Z$ a phase‑flip, and $Y$ a combined bit‑and‑phase flip.
This before‐round channel represents idling errors and any residual noise from previous operations.

After each Clifford operation in the syndrome-extraction circuit (including the entangling CNOTs), we apply an independent depolarizing error noise with probability $p$. In addition, we model measurement imperfections by applying a bit-flip with probability $p$ immediately before each measurement. Finally, before the start of every syndrome-measurement round, each data qubit undergoes an independent single-qubit depolarizing channel with probability $p$. 

In our simulation for throughput comparison results, we have considered $p=10^{-2}$ and $p=10^{-3}$, a regime that is expected to become accessible in the relatively near future. Furthermore, for logical error rate scaling results, we considered $p\in \{10^{-3},2 \times 10^{-3},4 \times 10^{-3},6 \times 10^{-3},8 \times 10^{-3},10^{-2}\}$.

Furthermore, we assume that each ancilla qubit is prepared perfectly in $\lvert0\rangle$ and measured perfectly in the $Z$-basis.  Any apparent syndrome‐readout error instead arises from prior depolarization propagating through the CNOT network.  Immediately after measurement, the ancilla are reset to $\lvert0\rangle$ noiselessly, ready for the next round.

Note that the Pauli-$X$ errors and Pauli-$Z$ errors are corrected independently. Therefore, in our model, we account for only Pauli‑$X$ failures by running  
memory $X$ experiment on every syndrome trajectory. In the memory‑$X$ run, the code is initialized in the logical $\lvert+\rangle_L$ state and undergoes $T_d$ rounds of syndrome extraction under circuit‑level depolarizing noise.  At the end of these rounds, we measure the logical $X$ operator. Any flip of this outcome indicates a logical‑$X$ error, whose probability we denote $P_{L}$. 
For the memory‑$Z$ run, the procedure is identical except that the code is initialized in $\lvert0\rangle_L$ and the final measurement of the logical $Z$ operator then yields the logical‑$Z$ failure.

We also assume ideal preparation of the logical $\lvert0_L\rangle$ state before syndrome extraction begins. If a shot runs to the final round $T_d$ without early abort, the decoder $\mathcal{D}(\bm X_{T_d})$ uses the full syndrome history $\bm X_{T_d}$
to apply a corrective Pauli and decide success or failure.  We treat the decoder’s correction as perfect, so any logical failure reflects an uncorrected Pauli error.


\section{Performance evaluations}
\label{sec:numerical_results}   
In this section, we present a comprehensive numerical study of three decoding schemes, FD, OSLA, and AdAbort, applied to both rotated surface and triangular color codes. For each type of code, we first compare the $\eta_\mathrm{dec}$ of each decoding scheme as a function of code distance for different physical error rates $p$. Then we look at the logical error rate scaling as a function of physical error rate $p$. Finally, we provide the sensitivity result of parameter $\theta$ with different physical error rates and code distances.

 \subsection{Intuition on expected scheme performance}
 \label{sec:intution}
Despite currently being the standard scheme, one can see that in practice, FD decoding tends to underperform adaptive schemes.  This is because the FD scheme always pays the full cost of all $T_d$ rounds, even when the syndrome history signals an almost certain success or failure after only a few rounds. On the other hand, OSLA improves on this by making a greedy decision at each round by comparing the immediate cost of stopping against the expected reduction in logical error from one more round. However, it remains short‐sighted beyond that single lookahead. In contrast, AdAbort learns the entire conditional failure probability, effectively incorporating information about all future rounds in one shot. As a result, in nearly every numerical result, one observes
$
 \text{FD} < \text{OSLA} < \text{AdAbort}
\quad\text{(w.r.t decoder efficiency).}
$
While these and other schemes can certainly be developed further, this work demonstrates the potential of adaptive abort to improve quantum computation.

Our experiments apply a circuit-level depolarizing noise model. Other QEC models and parameter regimes may differ in performance. Nonetheless, we anticipate similar conclusions: a well-chosen abort rule increases efficiency, by focusing computational resources and decoder effort on high-potential runs. Moreover, for larger codes, low-success runs appear to occur earlier in execution, leading to further benefits as code distance and computational complexity increase.

\subsection{Performance on surface codes}
In our surface code benchmarks, we focus on the rotated surface and employ an MWPM decoder. All syndrome extraction circuits were generated with \texttt{Stim} to obtain $T_d=d$ rounds of syndrome measurements per shot. 

\subsubsection{Decoder efficiency as a function of code distance}

\begin{figure}[htb]        
  \centering
  \begin{subfigure}{0.5\columnwidth}
    \includegraphics[width=\linewidth]{ 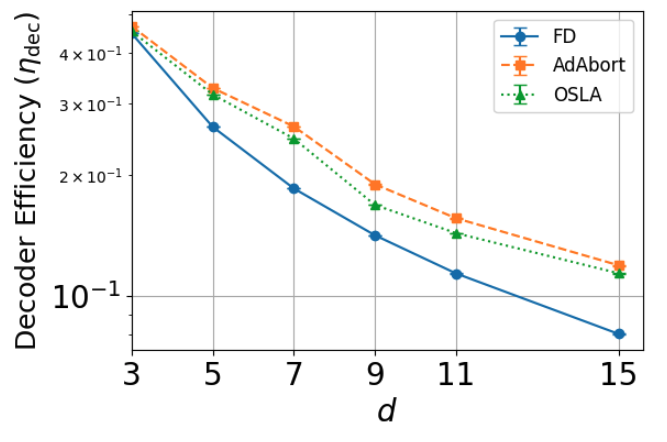}
    \caption{$p=10^{-2}$}
    \label{fig:variable_d_2}
  \end{subfigure}\hfill
  \begin{subfigure}{0.5\columnwidth}
    \includegraphics[width=\linewidth]{ 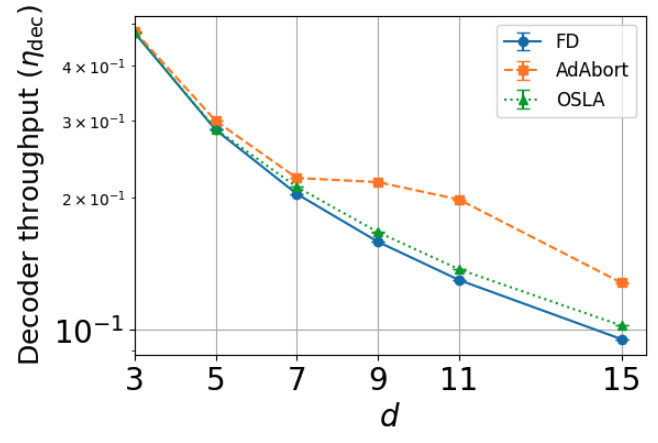}
    \caption{$p=10^{-3}$}
    \label{fig:variable_d_3}
  \end{subfigure}
  \caption{Decoder efficiency as a function of code distance $d$ for the surface code. Each data point runs for $2\times 10^5$ shots.}
  \label{fig:3}
\end{figure}

 The results from Figure~\ref{fig:variable_d_2} and Figure~\ref{fig:variable_d_3} confirm that early aborts by AdAbort decoding using learned failure probability yield substantial higher decoder efficiency, especially in the regime of moderate to large code distances.

 Figure \ref{fig:variable_d_2} shows the comparison of all three decoding schemes for $p=10^{-2}$. 
  All three curves fall roughly exponentially with $d$, reflecting the fact that measurement and decoding costs grow linearly in $d$. In Figure~\ref{fig:variable_d_3}, we repeat the same comparison at a lower physical error rate of $p=10^{-3}$. 
When the error rate is reduced, the efficiency rises across all three schemes and the decay with $d$ is more gradual. At small distances from $d=3$ to $d=5$, the three curves overlap, since the opportunity to abort early remains limited. As $d$ grows, AdAbort uses its ability to exploit the full syndrome measurement history when errors are less likely. OSLA still outperforms FD, but the relative gap between OSLA and AdAbort narrows slightly at this lower error regime. 
 
 It is clear from Figure~\ref{fig:variable_d_2} and Figure~\ref{fig:variable_d_3} that the FD decoding scheme has the lowest decoder efficiency, since it always pays for the complete $T_d$ rounds of syndrome measurements. OSLA improves on this by performing a one‑step lookahead at each round and aborting when the next‐round failure probability exceeds a decision threshold. AdAbort, in contrast, uses a neural network to estimate the logical‐error probability $\hat p_{\mathrm{err}}(\bm X_ t)$ and aborts as soon as that estimate crosses a tuned abort threshold, $\theta$. For each distance $d$, we performed a sweep over $\theta$ (as described in~\ref{subsub:optimal_threshold}) and selected the value that maximizes efficiency. As a result, AdAbort consistently achieves the highest decoder efficiency. 




At the higher noise rate $p=10^{-2}$ (Figure~\ref{fig:variable_d_2}), AdAbort improves decoder efficiency by about $4 \%$ at $d=3$, $25 \%$ at $d=5$, $42 \%$ at $d=7$, and nearly $50 \%$ by $d=15$ over the FD. Even at the lower noise rate $p=10^{-3}$ (Figure~\ref{fig:variable_d_3}), where errors are much rarer, it still recovers $1\%-9\%$  ($5\%$ at $d=5$) of latency for small and mid‑range distances, and up to $35\%$ at $d=15$ over the FD.



\subsubsection{Logical error rate scaling}
\begin{figure}[h!]
\centering
  \medskip
  \centering
  \includegraphics[width=10cm]{ 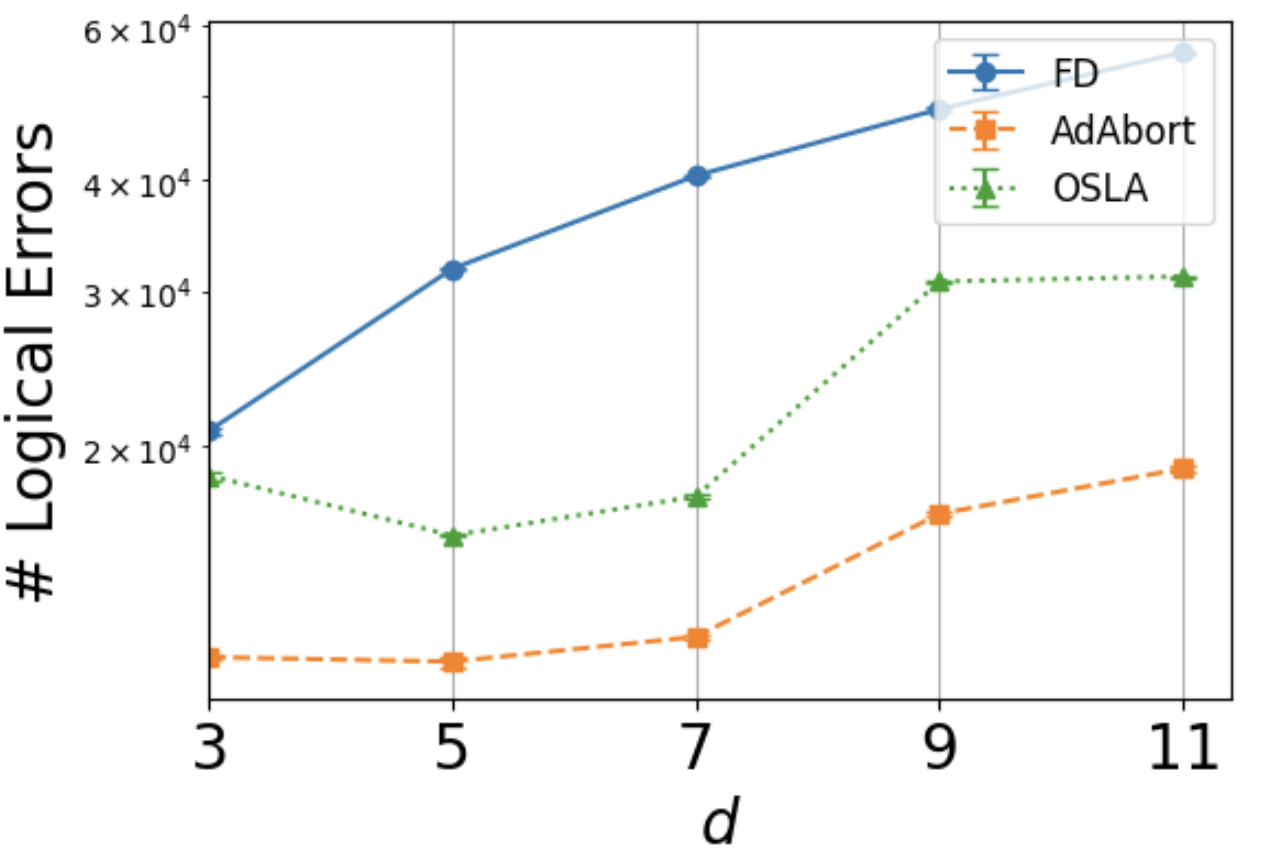}
\caption{Number of logical errors as a function of code distance for $p=10^{-2}$. Each point was obtained by running $5\times10^6$ shots.}
\label{fig:logical_failure}
\end{figure}

We also compare the number of logical errors as a function of code distance $d$ in Figure~\ref{fig:logical_failure}.  As $d$ increases, FD decoding performs the worst: its logical‐error count rises from roughly $4\times10^3$ at $d=3$ to over $10^4$ by $d=11$.  The OSLA and AdAbort schemes also see error rates that grow with $d$, but much more slowly than FD.  Across the entire distance range, AdAbort achieves the fewest logical failures.
\begin{figure}[h!]
\centering
  \medskip
  \centering
  \includegraphics[width=10cm]{ 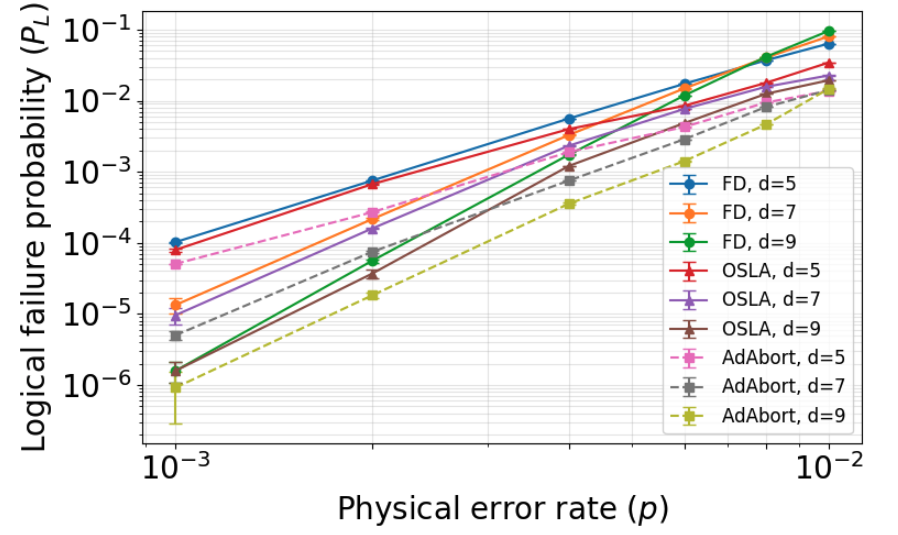}
\caption{For the three decoding schemes, in the physical error regime \(p=\{10^{-3},2 \times 10^{-3},4 \times 10^{-3},6 \times 10^{-3},8 \times 10^{-3},10^{-2}\}\) over $5\times10^7$ shots.
}
\label{fig:logical_error}
\end{figure}

Figure~\ref{fig:logical_error} shows the logical failure probability of a rotated surface code as a function of depolarizing error rate $p$, for code distances $d=5,7,9$ and three decoding strategies: FD, OSLA, and AdAbort. Each data point was obtained over $5\times10^7$ shots. All curves exhibit a clear monotonic increase of $P_{L}$ with $p$, and an exponential suppression of logical errors as $d$ grows. AdAbort decoding yields the lowest logical error rates at every $(d,p)$ point, followed by OSLA and then FD.

\subsubsection{Sensitivity of abort threshold}
\label{subsub:optimal_threshold}
\begin{figure}[htb]        
  \centering
  \begin{subfigure}{0.5\columnwidth}
    \includegraphics[width=\linewidth]{ 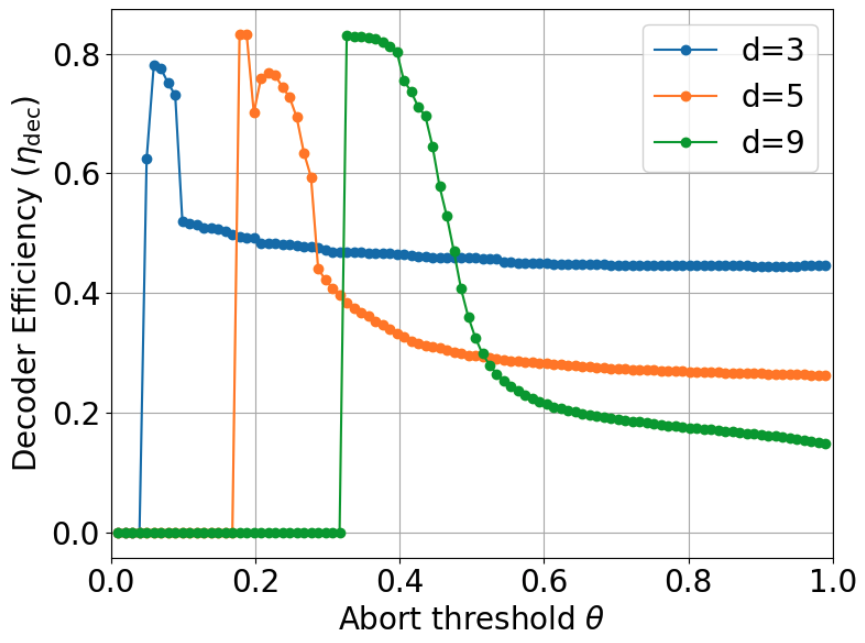}
    \caption{$p=10^{-2}$}
    \label{fig:optimal_theta}
  \end{subfigure}\hfill
  \begin{subfigure}{0.5\columnwidth}
    \includegraphics[width=\linewidth]{ 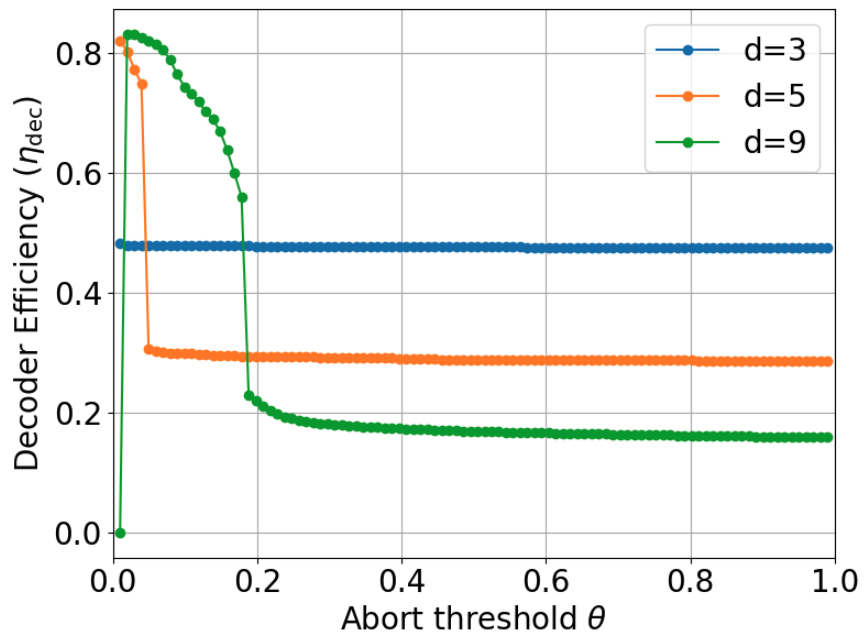}
    \caption{$p=10^{-3}$}
    \label{fig:optimal_theta2}
  \end{subfigure}
  \caption{Decoder efficiency of the AdAbort decoding scheme as a function of the abort threshold $\theta$ for rotated‑surface codes.}
  \label{fig:side-by-side-theta}
\end{figure}

When deploying the AdAbort decoding scheme with an abort module as shown in Figure~\ref{fig:abort_module}, the choice of the abort threshold $\theta$ is essential for its performance. It controls the trade‑off between saving syndrome measurement rounds and avoiding unnecessary aborts of correctable shots. In Figure~\ref{fig:optimal_theta} and Figure~\ref{fig:optimal_theta2}, we have plotted the decoder efficiency of the AdAbort decoding scheme as a function of $\theta$. We see that each code distance $d$ requires a distinct optimal window for $\theta$. Smaller codes, for example $d=3$, peak at low thresholds. This is because error events become obvious very early. On the other hand, larger codes $d=7$ require higher thresholds before the predictor is confident enough to abort.

For the small codes like $d=3$ and $d=5$, the efficiency jumps up and then stays high over a wide range of $\theta$. This indicates a robust choice of $\theta$ anywhere near the peak. In contrast, $d=7$ shows a narrower optimal $\theta$ window, so precise choice of $\theta$ becomes more critical as code size increases. 
Furthermore, at large $\theta$ values beyond the optimal window, efficiency degrades gently for smaller $d$ but more steeply for larger $d$. This shows that the cost of waiting too long, incurring many extra rounds before aborting, becomes more expensive when each round itself is costlier.

Comparing the two plots in Figures~\ref{fig:optimal_theta} and~\ref{fig:optimal_theta2} shows that lower physical error rates push the optimal $\theta$ window downwards and make it much narrower. This shows that the optimal $\theta$ window also depends on the circuit‑level depolarizing noise rate. At a higher error rate ($p=10^{-2}$), syndromes light up early, so the predicted logical failure probability will be high. Therefore, by choosing a mid‑range threshold, we can still achieve close‑to‐peak throughput over a broad range. In contrast, at a lower error rate ($p=10^{-3}$), failures are much rarer and the predicted $\hat p_t$ will be too low.  The only way to catch those few true failures is to set $\theta$ extremely low, and the throughput peak collapses into a very narrow range. Moreover, across both regimes, larger code distances require higher values of $\theta$.

\subsection{Performance on color codes}
For the triangular color code benchmarks, we generate syndrome extraction circuits with \texttt{Stim} as done in~\cite{lee2025color}, varying the code distance $d$ to produce $T_d=d$ rounds of check measurements per shot.  Each syndrome trajectory is then decoded by a concatenated MWPM~\cite{lee2025color} decoder. 
We show that early abort strategies yield higher decoder efficiency and higher logical‑error suppression on color codes, implying their broad applicability across topological QEC families.

Note that the performance order of decoding schemes FD, OSLA, and AdAbort will remain the same as that for the surface codes. Furthermore, the explanation for the different plots, such as efficiency as a function of code distance, logical error rate scaling and sensitivity of abort threshold, will be similar. Therefore, we provide the plot for the triangular color code with less explanation for brevity.

\begin{figure}[h!]
\centering
  \medskip
  \centering
  \includegraphics[width=10cm]{ 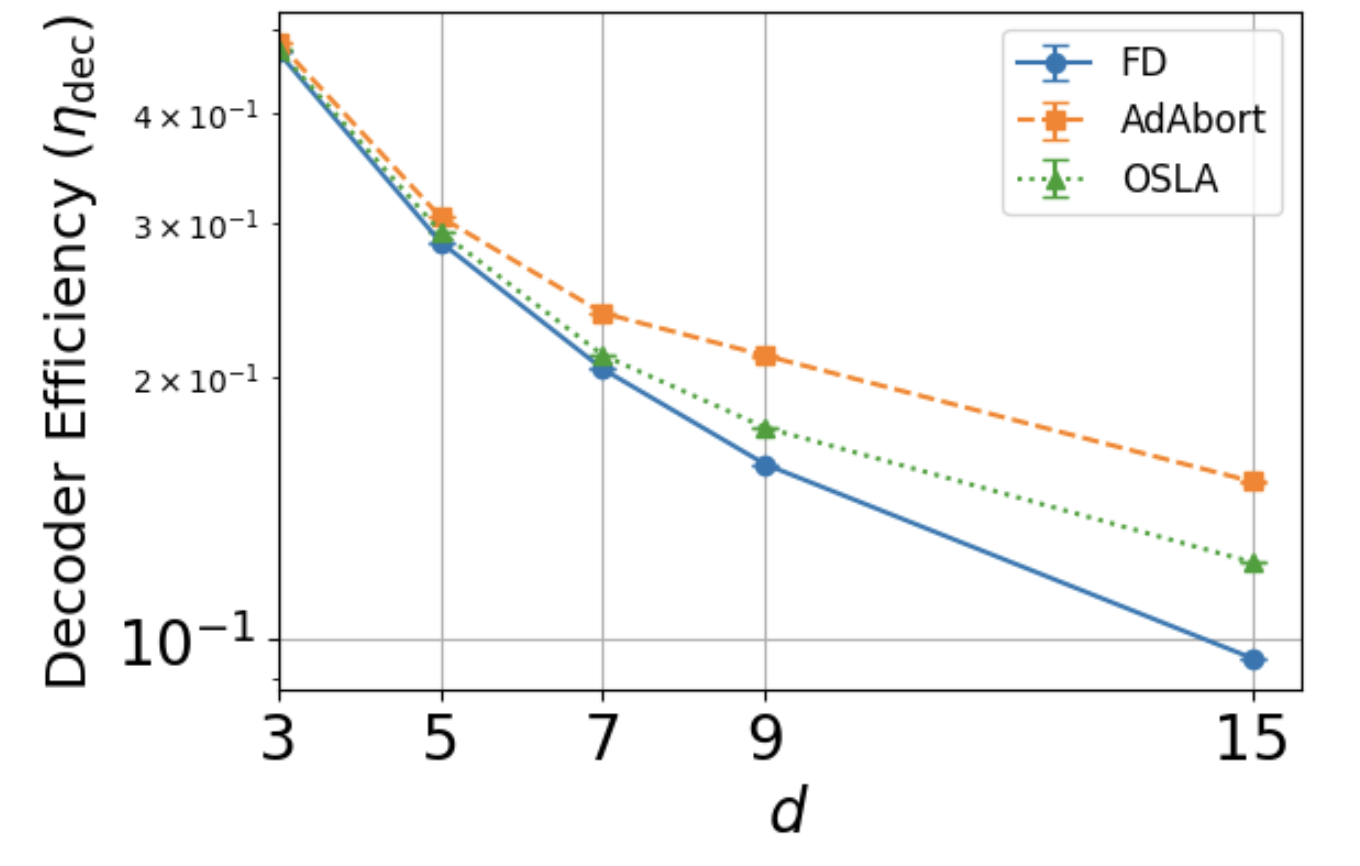}
\caption{Decoder efficiency as a function of code distance for the triangular color code, when $p=10^{-3}$.}
\label{fig:variable_d_colorcode}
\end{figure}

From Figure \ref{fig:variable_d_colorcode}, we see that AdAbort consistently outperforms FD decoding across all code distances for the triangular color code. For $p=10^{-3}$, even at the smallest distance $d=3$, AdAbort delivers an almost $3 \%$ decoder efficiency gain over FD; as the code grows, the benefit rises, reaching nearly $60 \%$ throughput gain by $d=15$. 

\begin{figure}[htb]        
  \centering
  \begin{subfigure}{0.48\columnwidth}
    \includegraphics[width=\linewidth]{ 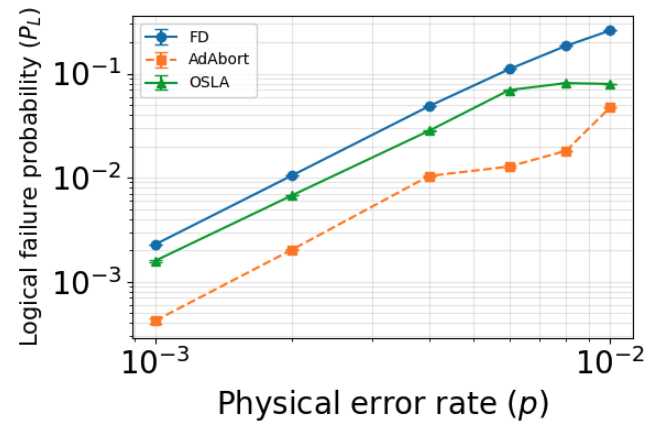}
    \caption{$d=5$}
    \label{fig:pl_5}
  \end{subfigure}\hfill
  \begin{subfigure}{0.48\columnwidth}
    \includegraphics[width=\linewidth]{ 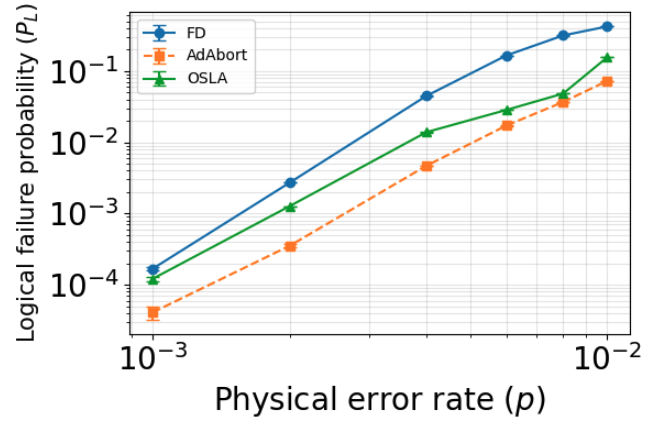}
    \caption{$d=9$}
    \label{fig:pl_7}
  \end{subfigure}
  \caption{$P_L$ as a function of $p$ for triangular color code with $d=\{5,9\}$. 
  The physical error regime \(p=\{10^{-3},2 \times 10^{-3},4 \times 10^{-3},6 \times 10^{-3},8 \times 10^{-3},10^{-2}\}\) 
  over $2\times10^7$ shots.
  }
  \label{fig:side-by-side-logical-error}
\end{figure}

 From Figure~\ref{fig:side-by-side-logical-error}, it is clear that all curves exhibit an apparent monotonic increase of $P_{L}$ with $p$, and an exponential suppression of logical errors as $d$ grows. AdAbort decoding yields the lowest logical error rates at every $(d,p)$ point, followed by OSLA and then FD. 
\begin{figure}[htb]        
  \centering
  \begin{subfigure}{0.47\columnwidth}
    \includegraphics[width=\linewidth]{ 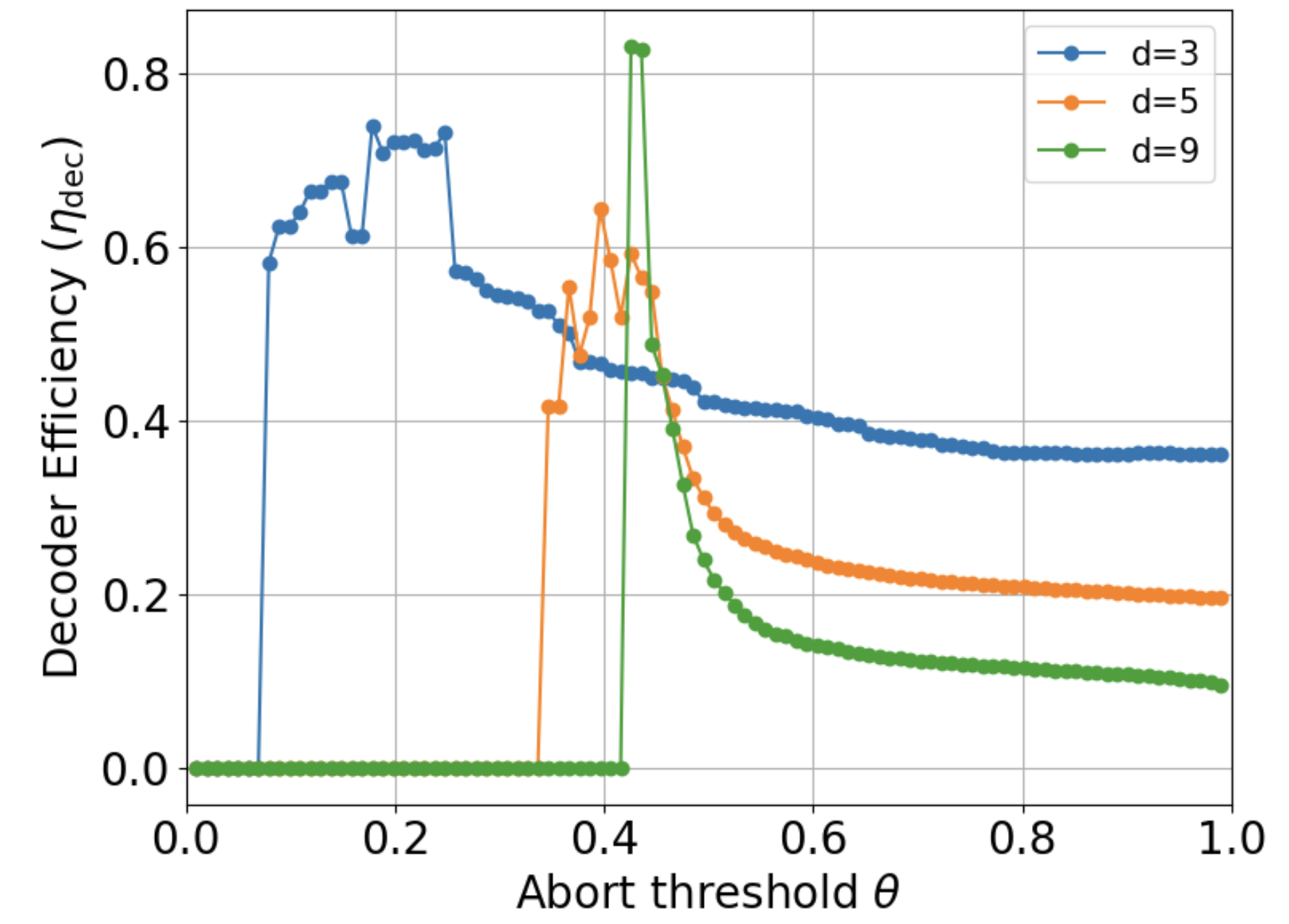}
    \caption{$p=10^{-2}$}
    \label{fig:optimal_theta_color}
  \end{subfigure}\hfill
  \begin{subfigure}{0.47\columnwidth}
    \includegraphics[width=\linewidth]{ 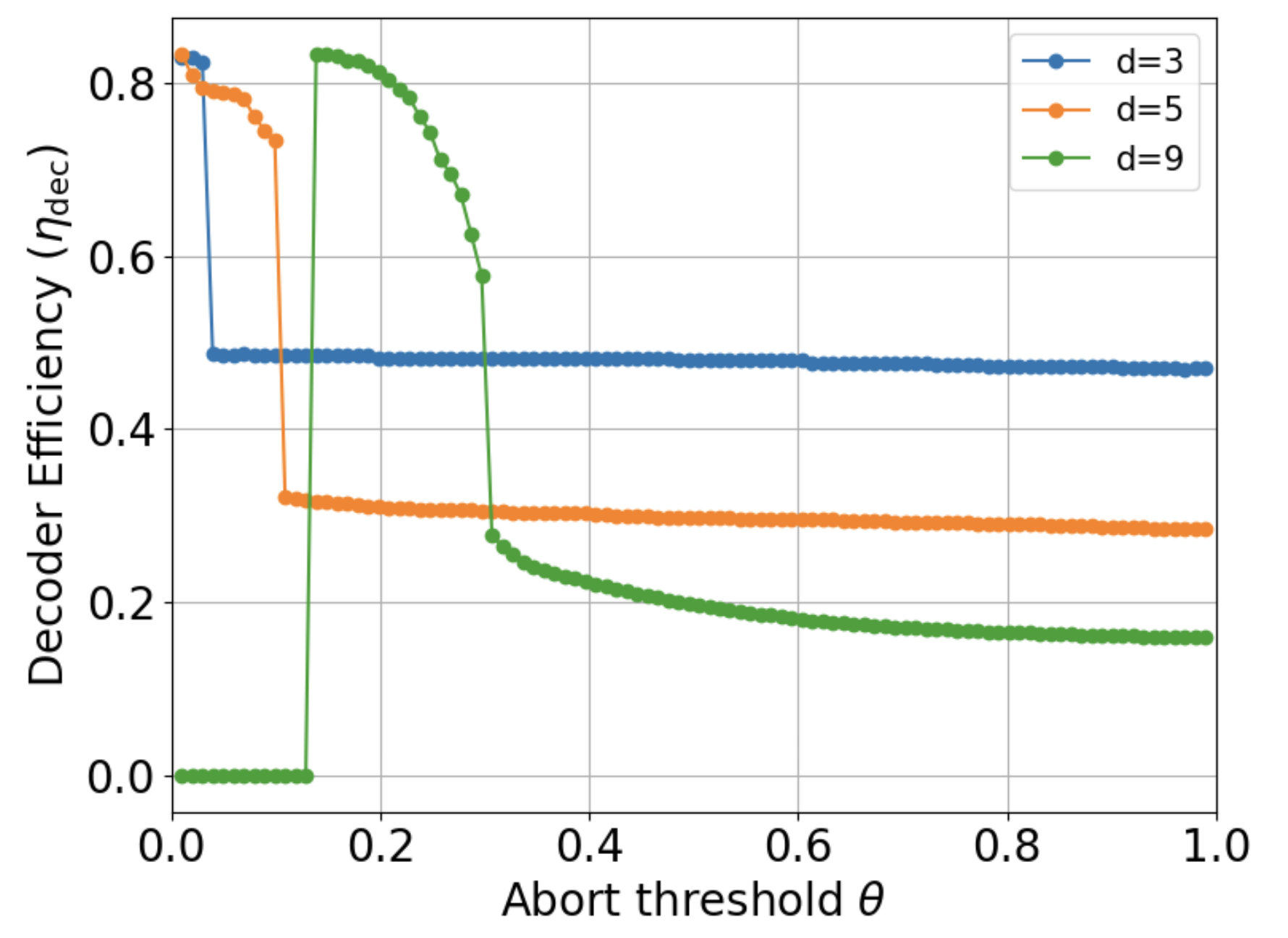}
    \caption{$p=10^{-3}$}
    \label{fig:optimal_theta_color2}
  \end{subfigure}
  \caption{Decoder efficiency as a function of the abort threshold $\theta$ for triangular color.}
  \label{fig:side-by-side-theta_color}
\end{figure}
In Figure \ref{fig:side-by-side-theta_color}, we plot the throughput of AdAbort as a function of the abort threshold $\theta$ for several code distances and two physical error rates. As with surface codes, the optimal $\theta$ window for the color code depends on both the noise rate $p$ and the code distance $d$. As $d$ increases, the peak throughput shifts to higher $\theta$ values, indicating that larger codes require a higher abort threshold.  Conversely, reducing the physical error rate $p$ shifts the optimal window to a lower $\theta$.


    \section{Conclusion}
\label{sec:colculsion}
In this work, we introduced the adaptive abort module to abort shots which are likely to have logical errors while decoding. Using this abort module, we have demonstrated the gain in decoder efficiency of FTQC operations across a wide range of code distances and error regimes. By tracking syndrome histories and applying either AdAbort or an OSLA decoding scheme, the abort module can terminate hopeless shots, saving precision quantum hardware execution time otherwise wasted on measurement and decoding time.  Our simulations on rotated surface and triangular color codes under realistic circuit-level depolarizing noise show the following: i) the decoder efficiency gains grow with code size: AdAbort delivers modest gains at small distances and exceeds a $35\%$ improvement for surface codes and an increase of almost $60\%$ for the color code over FD decoding by $d=15$, ii) these benefits come with minimal architectural impact—the abort module can be integrated onto existing QEC architecture without altering the quantum hardware.

Looking ahead, we expect further gains in decoder efficiency by improving the learned predictor used by AdAbort (e.g., better architectures, calibration, or training objectives) and by developing alternative adaptive schemes that more directly and efficiently predict eventual logical failure from partial syndrome information. Another important direction is to extend adaptive aborting beyond memory experiments to the execution of full fault-tolerant quantum algorithms, where early-termination and restart policies could interact with algorithmic structure. Finally, it will be interesting to study adaptive-abort strategies in distributed settings, such as modular or networked quantum architectures, where abort decisions may need to be coordinated across nodes and could reduce communication, synchronization, and decoding overheads.

\textbf{Acknowledgements:}
 This research was funded by the EPSRC funded INFORMED-AI project EP/Y028732/1.

\bibliographystyle{IEEEtran}
\bibliography{reference}

\end{document}